\documentclass[aps,pra,preprint,groupedaddress,showpacs,%
tightenlines,
nofootinbib,floatfix]{revtex4}

\usepackage{amsmath,amssymb,amsfonts,bm}
\usepackage{graphicx}
\usepackage{dcolumn}
\usepackage{mathrsfs} 

\renewcommand{\vec}[1]{\bm{\mathrm{#1}}} 

\begin{document}

\title{Resonant $\boldsymbol{dt\mu}$ formation in condensed hydrogens}

\author{Andrzej Adamczak}
\email{andrzej.adamczak@ifj.edu.pl}
\affiliation{Institute of Nuclear Physics, Polish Academy of Sciences,
Radzikowskiego 152, PL-31342~Krak\'ow, Poland}

\author{Mark P. Faifman}
\email{mark@rogova.ru}
\affiliation{Russian Research Center Kurchatov Institute, 
Kurchatov Square~1, RU-123182~Moscow,
Russia}

\date{\today}

\begin{abstract}
  Resonant formation of the muonic molecule $dt\mu$ in $t\mu$ atom
  collision with condensed H/D/T targets is considered. A~specific
  resonance correlation function, which is a~generalization of the Van
  Hove single-particle correlation function, is introduced to calculate
  the resonant-formation rate in such targets. This function is derived
  in the case of a~polycrystalline harmonic solid. Also is found
  a~general asymptotic form of the resonance correlation function for
  high momentum transfers, valid for any solid or dense-fluid
  hydrogen-isotope target.

  Numerical calculations of the rates are performed for solid molecular
  hydrogens at zero pressure, using the Debye model of an isotropic
  solid. It is shown that condensed-matter effects in resonant formation
  are strong, which explains some unexpected experimental results. In
  particular, the resonance profiles are affected by large zero-point
  vibrations of the hydrogen-isotope molecules bound in the considered
  crystals, even for high ($\sim$~1~eV) collision energies. This is
  important for explanation of the time-of-flight measurements of the
  $dt\mu$-formation rate, carried out at TRIUMF. The calculated mean
  values of the $dt\mu$-formation rate in solid D/T targets, for fixed
  target temperatures and steady-state conditions, are in good agreement
  with the PSI and RIKEN-RAL experiments.
\end{abstract}


\maketitle

\section{Introduction}

A~theoretical study of resonant formation of the muonic molecular ion
$dt\mu$ in condensed hydrogen-isotope targets is the main subject of
this paper. Formation of $dt\mu$ is a~key process of muon-catalyzed
fusion ($\mu$CF) in a~D/T mixture, which attracted particular interest
because one muon can catalyze more than 100 
fusions~\cite{gers77,jone86,breu87,acke99} according to the reaction
\begin{equation*}
  dt\mu \longrightarrow \, ^{4}\text{He}+n+\mu^{-}+\text{17.6~MeV} \,. 
\end{equation*}
Investigation of the $\mu$CF cycle in various hydrogen-isotope targets
is also important for studies of various phenomena in atomic, molecular,
and nuclear physics (see reviews~\cite{breu89,pono90,froe92}).

Resonant $dt\mu$ formation is due to the presence of the loosely bound
state of $dt\mu$~\cite{gers77} with the rotational quantum number~$J=1$,
the vibrational quantum number~$v=1$, and the binding energy
$\varepsilon_{Jv=11}\approx{}-0.63$~eV. Theoretical methods for
calculation of the resonant-formation rates were developed for
many years (see e.g., Refs.~\cite{vesm67,petr85,mens86b,mens88c,%
cohe89,faif89,faif91,leon94,petr96}). %
These methods, taking into account resonant formation in $t\mu$
collision with one or few molecules, give good agreement with the
experimental data for dilute gaseous targets. However, such theory is
unable to explain various phenomena found in experiments with dense
fluid and solid hydrogen-isotope targets. This concerns a~nonlinear
dependence of the formation rate on the target
density~\cite{acke99,aver01}, puzzling temperature
effects~\cite{kawa03}, and the resonance profiles determined by the
time-of-flight experiments~\cite{fuji99,fuji00,porc01,mars01}.
Therefore, it is necessary to consider the influence of many-body
effects on muonic-molecule formation. In particular, various collective
phenomena can significantly change this process, which one can expect
knowing their role in resonant neutron absorption by nuclei bound in
condensed matter~\cite{lamb39,sing60}.

Condensed-matter effects in resonant neutron absorption can be expressed
in terms of the single-particle correlation function~\cite{sing60}, which
has been introduced by Van~Hove~\cite{vanh54} for description of
incoherent neutron scattering. This function depends on energy and
momentum transfer to a~target and its properties. It is possible to
adapt this formalism to the case of resonant muonic-molecule formation.

First estimation of the $dt\mu$-formation rate in solid molecular
hydrogens was given by Fukushima~\cite{fuku93}. He employed
a~correlation-function formalism, performed ab initio calculation of
lattice dynamics to determine target properties, and demonstrated an
important role of phonon processes in resonant $dt\mu$ formation. His
calculation was limited to high target pressures ($\sim{}10$~kbar),
where solid hydrogens are classical crystals. However, in $\mu$CF
experiments, only zero or low pressures ($\ll{}10$~kbar) have been
applied. As a result, the solid-hydrogen targets are quantum crystals
with large amplitudes of zero-point vibrations of the molecules in the
lattices and very different properties. Thus, a~special approach is
necessary to solve lattice dynamics~\cite{silv80,soue86}. Owing to this
fact and to a~rough estimation of the transition-matrix elements for
$dt\mu$ formation, the results of Ref.~\cite{fuku93} are about five
times greater than the rates determined in the
experiments~\cite{jone86,acke99}. Moreover, the temperature dependence
of the calculated formation rate, for D$_2$ molecule bound in solid~D/T,
is opposite to what has been recently seen in the RIKEN-RAL
experiment~\cite{kawa03}.

A~theoretical method of calculating the resonant $dd\mu$-formation rate,
valid also for low-pressure solid hydrogens, has been presented in
detail in~Ref.~\cite{adam01}. The correlation function used for
description of properties of solid polycrystalline~D$_2$ has been derived
for the Debye model of an isotropic solid. The model parameters, such as
the Debye temperature and the lattice constants, has been taken from the
available data including quantum-crystal effects~\cite{silv80,soue86}.
Since the resonances in $dd\mu$ formation on an free D$_2$ molecule are
very narrow, their profiles are well-described by the delta function. As
a~result, the corresponding formation rates in a~solid are expressed in
terms of the same incoherent correlation function that is employed for
description of incoherent neutron scattering. The theoretical
$dd\mu$-formation rates lead to the time spectra of $dd$-fusion products
that are in good agreement with the data taken at~TRIUMF~\cite{know97}.

Below we present a~method of calculation of the
$dt\mu$-resonant-formation rates in condensed hydrogens, for wide
intervals of pressure and $t\mu$ collision energy. The profiles of
$dt\mu$ resonances for a~free-molecule are described by the Breit-Wigner
function~\cite{petr85,mens88}. In~Ref.~\cite{sing60}, such profiles have
been taken into account for neutron or $\gamma$-ray resonant absorption
by heavy nuclei. It has been assumed that the nuclear mass is not
practically changed after absorption. As a~result, a~standard incoherent
correlation function was sufficient for description of this process. In
the case of muonic molecule formation in hydrogens, the mass of a~target
molecule increases greatly after muonic-atom absorption and creation of
a~small muonic-molecular ion. Therefore, we introduce a~special
resonance correlation function that includes this effect into the
target dynamics. Only at lowest collision energies ($\lesssim{}10$~meV),
considered in~Refs.~\cite{fuku93,adam01}, an approximation that
neglects this mass change can be applied since then resonant formation
takes place practically in a~rigid lattice. Such approach is valid for
interpretation of experiments performed at lowest temperatures and
well-described by steady-state kinetics. On the other hand, correct
explanation of the time-of-flight experiments using energetic
($\sim{}1$~eV) beams of muonic atoms~\cite{fuji00,porc01,mars01} require
the knowledge of the formation rates at intermediate and higher
energies.

In Sec.~\ref{sec:freemol}, a~brief description of resonant $dt\mu$
formation in an~isolated hydrogen-isotope molecule is given. A~method of
calculation of the formation rates in condensed targets, using the
energy-dependent transition-matrix elements obtained for a~single
molecule, is discussed in~Sec.~\ref{sec:formalism}. In particular, the
formulas for the resonant-formation rates in harmonic polycrystalline
hydrogens are derived. They can be applied to both $dt\mu$ and $dd\mu$
resonant formation. The results of numerical calculations for $dt\mu$
formation in low-pressure solid hydrogens are presented
in~Sec.~\ref{sec:dtm-sol}. They have been obtained using a~full set of
the energy-dependent transition-matrix elements calculated for the free
molecules HD, D$_2$, and DT. The $dt\mu$ formation rates for some
typical solid targets are shown as functions of the $t\mu$ kinetic
energy and target temperature. In particular, contributions from
different resonances to the total formation rates and influence of the
ortho-D$_2$ and para-D$_2$ concentration in a~target on the formation
rates are considered. A~comparison of the calculated mean rates with
some experimental results is performed.

\section{Resonant formation in a~free molecule}
\label{sec:freemol}

First we consider resonant formation of $dt\mu$ (the reasoning is
analogical for the $dd\mu$ case) in the following reaction:
\begin{gather*}
  (t\mu)_F + (\text{D}C)^I_{\nu_iK_i} \longrightarrow 
  \bigl[(dt\mu)^S_{Jv}\,cee\bigr]_{\nu_fK_f} \,, \\
  C = \text{H, D, or T} \quad \text{and} \quad  
  c = p,\, d,\, \text{or~} t \,,
\end{gather*}
where D$C$ is a~free molecule in the initial rotational-vibrational
state $(\nu_iK_i)$ with total nuclear spin~$\vec{I}$. This spin is taken
into account for D$C$=D$_2$. The $t\mu$ atom has total spin $\vec{F}$
and center-of-mass (CMS) kinetic energy~$\varepsilon$. The molecular
complex $[(dt\mu)cee]$ is created in the rotational-vibrational
state~$(\nu_fK_f)$ and the molecular ion $dt\mu$, which plays the role
of a~heavy nucleus of the complex, has total spin~$\vec{S}$. This
process takes place due to the presence of a~loosely bound state of
$dt\mu$ with rotational number~$J=1$ and vibrational number~$v=1$. The
binding energy~$|\varepsilon_{Jv=11}|$ released in the reaction above is
transferred to rotational-vibrational degrees of freedom of the created
molecular complex $[(dt\mu)cee]$. The resonance condition is fulfilled
when $\varepsilon$ takes a~specific value~$\varepsilon^0_{if}$.  This is
so-called Vesman's mechanism of muonic-molecule formation, introduced
in~Ref.~\cite{vesm67} for the $dd\mu$ case. 
\begin{figure}[htb]
  \begin{center}
    \includegraphics[width=8cm]{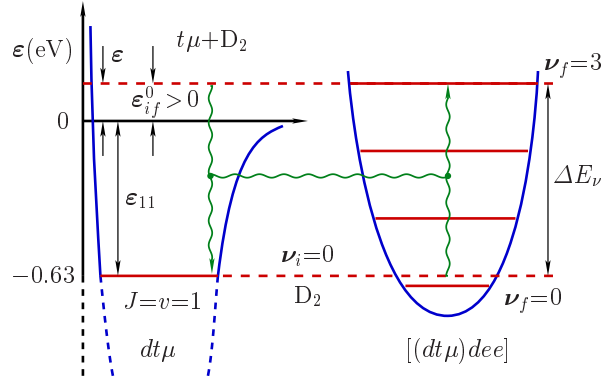}
    \caption{Energy diagram for Vesman's mechanism of resonant
      $dt\mu$ formation in collision of a~$t\mu$ atom with an isolated
      D$_2$ molecule.
    \label{fig:dtmbal_free}}
  \end{center}
\end{figure}
In Fig.~\ref{fig:dtmbal_free} is shown a~scheme of energy balance for
the $t\mu+$D$_2$ case. The formation rate
$\lambda^{SF}_{\nu_iK_i,\nu_fK_f}$ depends on the elastic width
$\varGamma^{SF}_{\nu_fK_f,\nu_iK_i}$ of $[(dt\mu)cee]$ 
decay~\cite{ostr80,padi88} through the channels:
\begin{equation*}
  \bigl[(dt\mu)^S_{Jv}\,dee\bigr]_{\nu_fK_f}
  \begin{cases}
    \xrightarrow[\varGamma^{SF}_{\nu_fK_f,\nu_iK_i}]
    &(t\mu)_F + (\text{D}C)^I_{\nu_iK_i} \\
    \xrightarrow[~~~~\lambda_f~~~~~]
    &\text{stabilization~processes,}
  \end{cases}
\end{equation*}
where $\lambda_f$ is the total rate of the stabilization processes,
i.e., deexcitations of $dt\mu$ and nuclear fusion in~$dt\mu$.
The value of $\varGamma^{SF}_{\nu_fK_f,\nu_iK_i}$ is given (in atomic
units $e=\hslash=m_e=$~1) by the equation
\begin{equation}
  \label{eq:el_width}
  \varGamma^{SF}_{\nu_fK_f,\nu_iK_i} = 2\pi A_{if} 
  \int \frac{d^3k}{(2\pi)^3} \, \lvert V_{if}(\varepsilon)\rvert^2
  \, \delta(\varepsilon^0_{if}-\varepsilon ) \,,
\end{equation}
The transition-matrix element is denoted by $V_{if}(\varepsilon)$ and
the resonance energy $\varepsilon^0_{if}$ is defined
in~Ref.~\cite{faif89}.  Factor~$A_{if}$ comes from averaging over
initial projections and summing over final projections of the spins and
angular momenta of the system. Vector~$\vec{k}$ is the momentum of
relative motion of the $t\mu$ atom and the molecule~D$C$, connected with
kinetic energy~$\varepsilon$ by the relation
\begin{equation*}
  \varepsilon = k^2/(2\mathcal{M}) \,,
\end{equation*}
in which $\mathcal{M}$ denotes the reduced mass of the system. The
general form of Eq.~(\ref{eq:el_width}) follows from the Fano theory of
resonant scattering~\cite{fano61}. Integration of this equation
over~$\vec{k}$ leads to
\begin{equation}
  \label{eq:gamma}
  \varGamma^{SF}_{\nu_fK_f,\nu_iK_i} = 
  \frac{\mathcal{M}\, k^0_{if}}{\pi} 
  A_{if}\lvert V_{if}(\varepsilon^0_{if})\rvert^2 \,,
  \quad k^0_{if} = k(\varepsilon^0_{if}) .
\end{equation}
In the Vesman model, the resonance width is very small, so that the
resonant-formation rate has the Dirac delta function profile
\begin{equation}
  \label{eq:resrate}
  \lambda^{SF}_{\nu_iK_i,\nu_fK_f} = 2\pi N_\text{mol}\,
  B_{if}\bigl|V_{if}(\varepsilon)\bigr|^2 
  \delta(\varepsilon-\varepsilon^0_{if}) \,.    
\end{equation}
where $N_\text{mol}$ is the density of hydrogen-isotope molecules in the
target. The coefficients $A_{if}$ and $B_{if}$ in
Eqs.~(\ref{eq:el_width}) and~(\ref{eq:resrate}) are defined below
\begin{equation}
  \label{eq:AB_def}
  \begin{split}
    A_{if} & = W_{SF}\, \xi(K_i) \,
    \frac{2K_i+1}{3\,(2K_f+1)} \, q_d  \,,  \\
    B_{if} & = W_{SF}\,
    \frac{2S+1}{3\,(2F+1)} \, q_d \,,
  \end{split}
\end{equation}
where $W_{SF}=1$ for~$dt\mu$ and 
\begin{equation*}
  W_{SF}  = 3\, (2F+1)
  \begin{Bmatrix}
    \tfrac{1}{2} & 1 & F \\
               1 & S & 1
  \end{Bmatrix}^{\!2}   
\end{equation*}
in the $dd\mu$ case. The curly brackets stand here for the Wigner
$3j$~symbol. For asymmetric molecules~D$C$, function~$\xi(K_i)=1$ and in
the case of D$_2$ we have
\begin{equation*}
  \xi(K_i)  = 
  \begin{cases}
    \frac{2}{3} & \text{ for } K_i \text{ even} \\
    \frac{1}{3} & \text{ for } K_i \text{ odd}  \,.
  \end{cases}
\end{equation*}
A~value of factor~$q_d$ is connected with the number of deuterons in
a~considered system. When $dt\mu$ is created in $t\mu$ collision with an
asymmetric molecule~D$C$, $q_d=1$, and if D$_2$ is a~target molecule,
$q_d=2$. In the case of $dd\mu$ formation in an asymmetric~D$C$ system,
factor~$q_d=2$. For $dd\mu$ formation in a~D$_2$ target, one has
$q_d=4$. Coefficient $B_{if}$ defined above differs from that introduced
in Ref.~\cite{faif89} since we omit here the Boltzmann factor describing
the population of the molecular rotational states in a~gas target. We
calculate the formation rate for a~fixed initial rotational state. This
rate is however averaged over total spin~$I$ of the target molecule. If
the muonic atoms in the target have a~steady kinetic-energy
distribution~$f(\varepsilon,T)$ at a~fixed target temperature~$T$,
Eq.~(\ref{eq:resrate}) can be additionally averaged over atomic
translational motion, which gives a~mean resonant
rate~$\tilde{\lambda}^{SF}_{\nu_iK_i,\nu_fK_f}(T)$.

Note that, for a~given set of the initial and final quantum numbers, the
resonance condition (cf.\ Fig.~\ref{fig:dtmbal_free})
\begin{equation}
  \label{eq:res_cond_free}
  \varepsilon = \varepsilon^0_{if}
\end{equation}
can be fulfilled only when the variable
$\varepsilon^0_{if}=\varepsilon_{11}+\varDelta{}E_{\nu}$ is positive.

In the case of resonant $dd\mu$ formation, the rates calculated using
Eq.~(\ref{eq:resrate}) agree very well with
experiments~\cite{peti99,peti01}. On the other hand, the assumption of
the delta-function profile for $dt\mu$ resonances has led to
inconsistency with experiments in gaseous D/T targets performed at low
temperatures~\cite{jone83,breu87,jeit95,acke99}. The measured rates are
much greater than the theoretical predictions based on the Vesman model.
It has been pointed by Petrov~\cite{petr85} that the $dt\mu$ resonances
should have broader Breit-Wigner profiles, owing to a~finite lifetime of
the complex. At low temperatures, this leads to significant
contributions to the formation rates~\cite{petr91} from the subthreshold
resonances $\varepsilon^0_{if}<0$. Thus, in a~general free-molecule
case, the resonance profile in Eq.~(\ref{eq:resrate}) can be described
by the Breit-Wigner function~\cite{petr85,mens88c}
\begin{equation}
  \label{eq:res_free_Breit}
  \lambda^{SF}_{\nu_iK_i,\nu_fK_f} = N_\text{mol}\, B_{if}
  \bigl|V_{if}(\varepsilon)\bigr|^2 \, \frac{\varGamma_S}
  {(\varepsilon-\varepsilon^0_{if})^2+\tfrac{1}{4}\varGamma_S^2} \,,
\end{equation}
where the total natural width~$\varGamma_S$ of the resonance is equal to
a~sum of the effective fusion rate~$\lambda_f$ and the total
rate~$\lambda^S_\text{bck}$ of back decay of the complex
\begin{equation}
  \label{eq:tot_gamma}
  \varGamma_S = \lambda_f + \lambda^S_\text{bck} \,.
\end{equation}
Equation~(\ref{eq:res_free_Breit}) was employed in
Refs.~\cite{petr94,petr96} for calculation of $dt\mu$ formation rate in
a~dilute D$_2$ gas, which led to agreement with the experimental
data~\cite{jeit95}. In the limit $\varGamma_S\to{}0$, the
rate~(\ref{eq:res_free_Breit}) tends to the Vesman
form~(\ref{eq:resrate}).

\section{Resonant formation in a~condensed target}
\label{sec:formalism}

\subsection{Method of calculation}
\label{sec:gen_formation}

When formation of a~muonic molecule takes place in a~dense target, it is
necessary to take into account interactions of the impinging muonic atom
with more than one molecule. In particular, energy transfer to many
molecules is possible, which results in a~quasiresonant character of the
formation process. A~quasiresonant mechanism of $dt\mu$ formation was
first considered in Ref.~\cite{mens86b}, for triple collisions
$t\mu+$D$_2+$D$_2$, in order to explain a~nonlinear density dependence
of the $dt\mu$ formation rate. In this case, formation is possible even
if the resonance condition~(\ref{eq:res_cond_free}) is not strictly
fulfilled, because an energy excess in the $t\mu+$D$_2$ system is
transferred to the second D$_2$ molecule. The three-body reactions and
broadening of the resonance profiles were then discussed in
Refs.~\cite{mens88,cohe89,petr91,leon94}. If a~target is condensed, it
is indispensable to take into account collective motions of target
molecules in the process of resonant formation.
\begin{figure}[htb]
  \begin{center}
    \includegraphics[width=8cm]{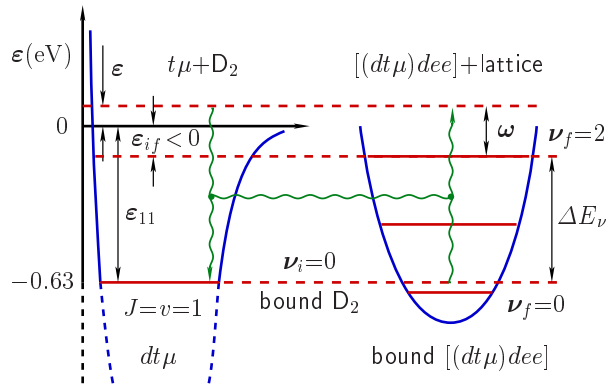}
    \caption{Energy diagram for quasiresonant formation of 
      $dt\mu$ in a~D$_2$ molecule bound in a~condensed target.
      \label{fig:dtmbal_sol}}
  \end{center}
\end{figure}
In~Fig.~\ref{fig:dtmbal_sol} is presented a~scheme of quasiresonant
$dt\mu$ formation in $t\mu$ collision with a~bulk condensed D$_2$
target. The energy balance, including energy transfer~$\omega$ to the
target, is shown for the subthreshold resonance corresponding to the
transition $\nu_i=0\to\nu_f=2$. Since the target molecule and the
complex $[(dt\mu)dee]$ are bound, the corresponding resonance
energy~$\varepsilon_{if}$ is different from the ``free'' resonant
energy~$\varepsilon^0_{if}$, characterized by the same set of the
quantum numbers.

Owing to a~certain analogy between resonant absorption of neutrons and
resonant formation of muonic hydrogen molecules, the methods developed
in neutron physics can be adapted for calculation of the rates of
resonant $dd\mu$ and $dt\mu$ formation. Resonant neutron absorption and
emission in condensed targets was first considered by
Lamb~\cite{lamb39}. His method was then generalized by Singwi and
Sj{\"o}lander~\cite{sing60}, using the single-particle response
function~$\mathcal{S}_i$~\cite{vanh54}, and applied for description of
resonant absorption and emission of $\gamma$~ray and neutrons in
condensed matter. In this Section are derived some expressions for the
rate of muonic molecule formation in muonic atom collision with
molecule~D$C$, bound in a~bulk hydrogen-isotope target.

\begin{figure}[htb]
  \begin{center}
    \includegraphics[width=5cm]{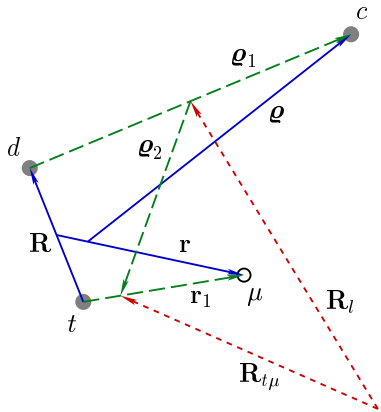}
    \caption{System of coordinates used for calculation of the
      formation rate of the complex~$[(dt\mu)cee]$ in a~condensed
      target.
    \label{fig:dtm_cmplx}}
  \end{center}
\end{figure}

A~Hamiltonian~$\mathscr{H}_\text{tot}$ of the system, consisting of
a~$t\mu$ atom in the 1$S$ state and a~bulk condensed D$C$~target, can be
written down as follows
\begin{equation}
  \label{eq:H_tot}
  \begin{split}
    \mathscr{H}_\text{tot} = \,   
    &\frac{1}{2M_{a\mu}} \boldsymbol{\nabla}^2_{R_{t\mu}}
    + \mathscr{H}_{t\mu}(\vec{r}_1) 
    + \mathscr{H}_{\text{D}C}(\vec{\varrho}_1) \\ 
    &+ V(\vec{r}_1,\vec{\varrho}_1,\vec{\varrho}_2) + \mathscr{H}  \,,
  \end{split}
\end{equation}
where $M_{a\mu}$ is the muonic atom mass and $\vec{R}_{t\mu}$ denotes
the position of $t\mu$ center of mass in the coordinate frame connected
with the target (see~Fig.~\ref{fig:dtm_cmplx}).
Operator $\mathscr{H}_{t\mu}$ is the Hamiltonian of a~free $t\mu$ atom,
$\vec{r}_1$ is the $t\mu$ internal vector, and $\mathscr{H}_{\text{D}C}$
denotes the internal Hamiltonian of a~free D$_2$ molecule. It is assumed
that $dt\mu$ formation takes place in $t\mu$ collision with the $l$th
molecule~D$C$. The position of its mass center in the target frame is
denoted by~$\vec{R}_l$; $\vec{\varrho}_1$~is a~vector connecting the
nuclei inside this molecule. Function~$V$~stands for the potential of
the $t\mu$--D$C$ interaction~\cite{faif89} that leads to resonant
$dt\mu$ formation. Vector $\vec{\varrho}_2$ connects the $t\mu$ and the
D$C$ centers of masses. We neglect contributions to potential~$V$ from
the molecules other than the $l$th molecule because we assume here that
distances between different molecules in the target are much greater
than the D$C$-molecule size. This assumption is valid for condensed
hydrogens under low pressure~\cite{silv80,soue86}. The kinetic
energy~$\varepsilon$ of the impinging muonic atom and its
momentum~$\vec{k}$ in the target frame are connected by the relation
\begin{equation*}
  \varepsilon=k^2/(2M_{a\mu}) \,.
\end{equation*}

\begin{figure}[htb]
  \begin{center}
    \includegraphics[width=6.5cm]{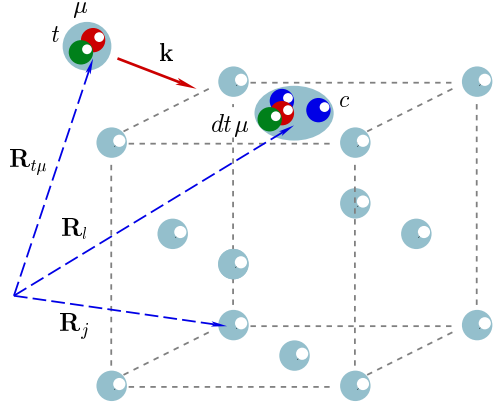}
    \caption{Position of the impinging $t\mu$ atom with respect 
      to the condensed target.}
    \label{fig:dtm_lattice}
  \end{center}
\end{figure}

The initial Hamiltonian~$\mathscr{H}$ of the condensed hydrogen-isotope
target, corresponding to the initial target energy~$E_0$, has the
following form: 
\begin{equation}
  \label{eq:hamilt0}
  \mathscr{H} = \sum_{j} \frac{1}{2M_j} 
  \boldsymbol{\nabla}^2_{R_j} +\sum_{j}\sum_{j'\neq j} U_{jj'} \,,
\end{equation}
where $\vec{R}_j$ is the position of the $j$th-molecule CMS (see
Fig.~\ref{fig:dtm_lattice}), $U_{jj'}$~denotes interaction between the
$j$th and $j'$th molecule, and $M_j$ is the mass of the $j$th molecule.

The coordinate part~$\varPsi_{\text{tot}}$ of the initial wave
function of the system can be written as a~product
\begin{equation}
  \label{eq:wf_tot1}
  \varPsi_{\text{tot}} = \psi_{t\mu}^{1S}(\vec{r}_1)\, 
  \psi_{\text{D}C}^{\nu_iK_i}(\vec{\varrho_1}) \,
  \exp(i\vec{k}\cdot\vec{R}_{t\mu})\, |0\rangle \,,
\end{equation}
where $|0\rangle$ stands for the initial wave function of the condensed
target, corresponding to the total energy~$E_0$. The eigenfunctions of
the operators $H_{t\mu}$ and $H_{\text{D}C}$ are denoted by
$\psi_{t\mu}^{1S}$ and~$\psi_{\text{D}C}^{\nu_iK_i}$, respectively.
Using the relation $\vec{R}_{t\mu}=\vec{R}_l+\vec{\varrho}_2$, the wave
function $\varPsi_{\text{tot}}$ takes the form
\begin{equation}
  \label{eq:wf_tot2}
  \varPsi_{\text{tot}} = \psi_{t\mu}^{1S}(\vec{r}_1)\, 
  \psi_{\text{D}C}^{\nu_iK_i}(\vec{\varrho_1}) \,
  \exp(i\vec{k}\cdot\vec{\varrho}_2)\, \exp(i\vec{k}\cdot\vec{R}_l) \,
  |0 \rangle \,,
\end{equation} 
which is similar to that used in the case of $dt\mu$ formation in
a~single molecule~D$C$, except the factor
$\exp(i\vec{k}\cdot\vec{R}_l)\,|0\rangle$. This factor depends only on
the CMS positions of the target molecules.

After formation of the $[(dt\mu)cee]$ complex, the total Hamiltonian of
the system is well approximated by the 
operator~$\mathscr{H}'_{\text{tot}}$
\begin{equation}
  \label{eq:H'_tot}
  \mathscr{H}_{\text{tot}}\approx 
  \mathscr{H}'_{\text{tot}} =  \mathscr{H}_{dt\mu}(\vec{r},\vec{R}) 
  + \mathscr{H}_\text{cplx}(\vec{\varrho}) 
  + V(\vec{\varrho},\vec{r},\vec{R}) 
  + \widetilde{\mathscr{H}} \,,
\end{equation}
where $\mathscr{H}_{dt\mu}$ is the internal Hamiltonian of the $dt\mu$
and vectors $\vec{r}$ and $\vec{R}$ denote its Jacobi coordinates.
Relative motion of the $dt\mu$ and nucleus~$c$ in the complex is
described by a~Hamiltonian~$\mathscr{H}_\text{cplx}$, which depends on
the corresponding internal vector~$\vec{\varrho}$. The final Hamiltonian
$\widetilde{\mathscr{H}}$ of the target is
\begin{equation}
  \label{eq:hamiltn}
  \begin{split}
    \widetilde{\mathscr{H}} &= \frac{1}{2M_\text{cplx}}
    \boldsymbol{\nabla}^2_{R_l} 
    +\sum_{j\neq l} \frac{1}{2M_j}\boldsymbol{\nabla}^2_{R_j}
    +\sum_{j}\sum_{j'\neq j} U_{jj'} \\
    &= \mathscr{H} + \varDelta\mathscr{H}\,,
  \end{split}
\end{equation} 
where
\begin{equation}
  \label{eq:deltaH}
  \varDelta\mathscr{H} = -\alpha\,\frac{1}{2M_{\text{D}C}} 
  \boldsymbol{\nabla}^2_{R_l} \,, \qquad  \alpha \equiv 
  1-\frac{M_{\text{D}C}}{M_\text{cplx}} \lesssim \frac{1}{2}\,,
\end{equation}
$M_\text{cplx}$ is the mass of the complex, and $M_{\text{D}C}=M_l$ is
the mass of the D$C$~molecule. A~small perturbation of potential~$V$,
due to replacement of the D$C$ center of mass by that of the complex, is
neglected here. The eigenfunction and eigenvalue
of~$\widetilde{\mathscr{H}}$ are denoted by $|\widetilde{n}\rangle$
and~$\widetilde{E}_n$, respectively. The coordinate
part~$\varPsi'_{\text{tot}}$ of the final total wave function of the
system is
\begin{equation}
  \label{eq:wf'_tot}
  \varPsi'_{\text{tot}}= \psi_{dt\mu}^{Jv}(\vec{r},\vec{R}) \, 
  \psi_\text{cplx}^{\nu_fK_f}(\vec{\varrho}) \, |\widetilde{n}\rangle \,, 
\end{equation}
where $\psi_{dt\mu}^{Jv}$ and $\psi_\text{cplx}^{\nu_fK_f}$ stand for
eigenfunctions of the corresponding Hamiltonians $H_{dt\mu}$
and~$H_\text{cplx}$.

For the initial $|0\rangle$ and final $|\widetilde{n}\rangle$ target
states and for a~fixed $t\mu$ spin~$F$, the energy-dependent
$dt\mu$-formation rate $\lambda^{SF}_{\nu_iK_i,\nu_fK_f}(\varepsilon)$
is calculated using the equation
\begin{equation}
  \label{eq:resratsol0}
  \begin{split}
    \lambda^{SF}_{\nu_iK_i,\nu_fK_f} =& \,
    N_\text{mol}\, B_{if}\, \lvert \mathcal{A}_{i0,fn} \rvert^2 \\
    &\times \frac{\varGamma_S}
    {(\varepsilon+E_0-\varepsilon^0_{if}-\widetilde{E}_n)^2 +
      \tfrac{1}{4}\varGamma_S^2} \,,
  \end{split}
\end{equation}
with the resonance condition
\begin{equation}
  \label{eq:res_cond}
  \varepsilon + E_0 = \varepsilon^0_{if} + \widetilde{E}_n \,.
\end{equation}
Equation~(\ref{eq:resratsol0}) is analogical to the Breit-Wigner
form~(\ref{eq:res_free_Breit}) used for a~free molecule. However, the
transition-matrix element is now given by
\begin{equation}
  \label{eq:matrixel1}
  \mathcal{A}_{i0,fn} =  \langle \varPsi'_{\text{tot}}\lvert  
  V \rvert \varPsi_{\text{tot}} \rangle \,.
\end{equation}
By virtue of Eqs. (\ref{eq:wf_tot2}) and (\ref{eq:wf'_tot}), the matrix
element~(\ref{eq:matrixel1}) can be written as a~product
\begin{equation}
  \label{eq:matrixel2}
  \mathcal{A}_{i0,fn} = \langle \widetilde{n}|\exp(i\vec{k}\cdot
  \vec{R}_l)|0\rangle \, V_{if}(\varepsilon) \,,
\end{equation}
where $V_{if}(\varepsilon)$ is the energy-dependent transition-matrix
element calculated for a~single molecule~\cite{faif89}. Averaging the
rate~(\ref{eq:resratsol0}) over a~distribution~$\rho_{n_0}$ of the
initial target states at a~given temperature~$T$ and summing over the
final target states leads to
\begin{equation*}
  \begin{split}
    \lambda^{SF}_{\nu_iK_i,\nu_fK_f} =& \,
    N_\text{mol}\, B_{if}\, |V_{if}|^2 \, \varGamma_S \\
    &\times \sum_{n,n_0}\, \rho_{n_0}\,
    \frac{\vert\langle \widetilde{n}\lvert
      \exp(i{\vec k}\cdot \vec{R}_l)|0\rangle\rvert^2}
    {(\varepsilon+E_0-\varepsilon^0_{if}-\widetilde{E}_n)^2 +
      \tfrac{1}{4}\varGamma_S^2} \; .
  \end{split}  
\end{equation*}
Factor $B_{if}$ is due to averaging over the initial projections and
summation over the final spin projections and over the
rotational-vibrational quantum numbers. The equation above
can be written down in the integral form
\begin{equation*}
  \begin{split}
    \lambda^{SF}_{\nu_iK_i,\nu_fK_f} =& \, 
    N_\text{mol}\, B_{if}\, |V_{if}|^2 \, \varGamma_S \\
    &\times \sum_{n,n_0}\, \rho_{n_0}\, 
    \vert\langle \widetilde{n}\lvert
    \exp(i{\vec k}\cdot \vec{R}_l)|0\rangle\rvert^2 \\
    &\times \int_{-\infty}^{\infty} dE\,
    \frac{\delta(E-\widetilde{E}_n+E_0)}
    {(\varepsilon-\varepsilon^0_{if}-E)^2 +
      \tfrac{1}{4}\varGamma_S^2} \; .  
  \end{split}
\end{equation*}
\begin{widetext}
  Now we introduce a~time variable~$t$ to eliminate the
  $\delta$~function in the equation above and then we involve
  time-dependent operators, which is familiar in scattering
  theory~\cite{akhi47,wick54}. Using the Fourier expansion of the
  $\delta$~function one obtains
\begin{equation*}
  \begin{split}
    \lambda^{SF}_{\nu_iK_i,\nu_fK_f} = \frac{1}{2\pi}
    N_\text{mol}\, B_{if}\, |V_{if}|^2 \, \varGamma_S
    \int_{-\infty}^{\infty} dt\, 
    \sum_{n,n_0} \, & \rho_{n_0} \,
    \vert\langle \widetilde{n}\lvert
    \exp(i{\vec k}\cdot \vec{R}_l)|0\rangle\rvert^2 
    \exp[it(\widetilde{E}_n-E_0)] \\
    &\times \int_{-\infty}^{\infty} dE \,\frac{\exp(-iEt)}
    {(\varepsilon-\varepsilon^0_{if}-E)^2 +
      \tfrac{1}{4}\varGamma_S^2}\,, 
  \end{split}
\end{equation*}
which, after integration over~$E$, gives
\begin{equation}
  \label{eq:resratsol2}
  \begin{split}
    \lambda^{SF}_{\nu_iK_i,\nu_fK_f} = N_\text{mol}\,
    B_{if}\,|V_{if}|^2  &\int_{-\infty}^{\infty} dt\, 
    \exp\left[-it\left(\varepsilon-\varepsilon^0_{if}\right) 
      -\tfrac{1}{2}\varGamma_S|t| \, \right] 
    \sum_{n,n_0} \rho_{n_0} \\
    &\times \langle 0\vert\exp(-i\vec{k}\cdot\vec{R}_l)
    \vert\widetilde{n}\rangle 
    \langle\widetilde{n}\vert\exp(it\widetilde{E}_n)
    \exp(i\vec{k}\cdot\vec{R}_l)\exp(-itE_0)\vert 0\rangle \,.
  \end{split}
\end{equation}
The matrix element in Eq.~(\ref{eq:resratsol2}) can be expressed as
follows
\begin{equation*}
  \begin{split}
    \langle\widetilde{n}\vert \exp(it\widetilde{E}_n) 
    \exp(& i\vec{k}\cdot\vec{R}_l)\exp(-itE_0)\vert 0\rangle 
    = \langle \widetilde{n}\vert \exp(it\widetilde{\mathscr{H}}) 
    \exp(i\vec{k}\cdot\vec{R}_l) 
    \exp(-it\mathscr{H})\vert 0\rangle \\
    &= \langle \widetilde{n}\vert\exp(it\widetilde{\mathscr{H}})
    \exp(-it\mathscr{H}) \exp(it\mathscr{H}) 
    \exp(i\vec{k}\cdot\vec{R}_l) 
    \exp(-it\mathscr{H})\vert 0\rangle \\ 
    &= \langle \widetilde{n}\vert\exp(it\widetilde{\mathscr{H}})
    \exp(-it\mathscr{H}) \exp[i\vec{k}\cdot\vec{R}_l(t)] 
    \vert{}0\rangle  \,,
  \end{split}
\end{equation*}
\end{widetext}
where $\vec{R}_l(t)$ denotes the Heisenberg operator
\begin{equation*}
  \vec{R}_l(t) = \exp(it\mathscr{H})\,\vec{R}_l\,
  \exp(-it\mathscr{H}) \,,
\end{equation*}
defined for all~$l$ and $t$.

Employing the identity %
$\sum_n |\widetilde{n}\rangle\langle\widetilde{n}|=1$
in~Eq.~(\ref{eq:resratsol2}) we obtain
\begin{equation}
  \label{eq:resratsol_gen}
  \lambda^{SF}_{\nu_iK_i,\nu_fK_f} = 
  2\pi N_\text{mol} B_{if} |V_{if}|^2 \, 
  \mathcal{S}_\text{res}(\vec{k},\varepsilon-\varepsilon_{if}^0) \,,
\end{equation}
where $\mathcal{S}_\text{res}$ is the %
\textit{resonance response function}
\begin{equation}
  \label{eq:S_res}
  \begin{split}
    \mathcal{S}_\text{res}(\vec{k},\varepsilon-\varepsilon_{if}^0) 
    \equiv \frac{1}{2\pi} \int_{-\infty}^{\infty} dt\,
    &\exp\bigl[-it(\varepsilon-\varepsilon_{if}^0) 
    -\tfrac{1}{2}\varGamma_S|t| \bigr] \\  
    &\times \mathscr{Y}_\text{res}(\vec{k},t) \,.
  \end{split}
\end{equation}
$\mathscr{Y}_\text{res}(\vec{k},t)$ denotes here  
the \textit{resonance correlation function} defined below
\begin{equation}
  \label{eq:func_Y}
  \begin{split}
    \mathscr{Y}_\text{res}(\vec{k},t) \equiv  
    \bigl\langle &\exp[-i\vec{k}\cdot\vec{R}_l(0)] 
    \exp(it\widetilde{\mathscr{H}})\\ 
    &\times \exp(-it\mathscr{H}) 
    \exp[i\vec{k}\cdot\vec{R}_l(t)]\bigr\rangle_T \,,
  \end{split}
\end{equation}
where $\langle\cdots\rangle_T$ stands for the quantum-mechanical
and statistical averaging at temperature~$T$.

On substitution $\widetilde{\mathscr{H}}=\mathscr{H}$ and
$\varGamma_S=0$ in the equations above, we recover the well-known
incoherent response function~$\mathcal{S}_\text{res}=\mathcal{S}_i$,
which describes incoherent neutron scattering in condensed
matter~\cite{vanh54,love84}. The approximation
$\widetilde{\mathscr{H}}=\mathscr{H}$ is valid when the mass of an
absorbed particle is much smaller than the mass of a~target atom or
a~molecule. This is a~common and good approximation when neutron
absorption by a~much heavier nucleus is considered. However, the
difference~$\varDelta\mathscr{H}$ between the Hamiltonians
$\widetilde{\mathscr{H}}$ and~$\mathscr{H}$ cannot be neglected in the
case of muonic molecule formation since the mass of muonic hydrogen atom
is comparable with that of hydrogen isotope molecule.

The partial width~$\varGamma^{SF'}_{\nu_fK_f,\nu_iK_i}$ of back decay of
the complex, bound in a~condensed target, is given by the expression
analogical to Eq.~(\ref{eq:el_width})
\begin{equation}
  \label{eq:part_back1}
  \varGamma^{SF'}_{\nu_fK_f,\nu_iK_i} = 2\pi A_{if} 
  \int \frac{d^{\,3}k}{(2\pi)^3} \, \lvert\,\mathcal{A}_{i0,fn}\rvert^2
  \, \delta(\varepsilon^0_{if} + \widetilde{E}_n 
  -\varepsilon -E_0) \,.
\end{equation}
Using the Fourier expansion of the $\delta$~function and proceeding
as in the case of quasiresonant formation process we obtain 
\begin{equation}
  \label{eq:elwidth_gen}
  \varGamma^{SF'}_{\nu_fK_f,\nu_iK_i} = 2\pi A_{if} 
  \int \frac{d^3k}{(2\pi)^3} \,\lvert V_{if}(\varepsilon)\rvert^2
  \, \widetilde{\mathcal{S}}_\text{res}(\vec{k},\varepsilon_{if}^0 
  - \varepsilon) \,, 
\end{equation}
where $\widetilde{\mathcal{S}}_\text{res}$ denotes
function~(\ref{eq:S_res}) calculated for the initial 
state~$\vert\widetilde{n}\rangle$, with~$\varGamma_S$ set to~zero.

In order to compare the calculated formation rates with experiments,  
the summed formation rates~$\lambda^F_{K_i}(\varepsilon)$ are introduced
\begin{equation}
  \label{eq:rate_sum}
  \lambda^F_{K_i} =  \sum_{\nu_f,K_f, S} 
  \lambda^{SF}_{\nu_i K_i \nu_f K_f} \,, \qquad \nu_i=0 \,. 
\end{equation}
In simulations of muon-catalyzed fusion involving energy-dependent rates
of various processes the ``absolute'' formation
rates~(\ref{eq:rate_sum}) should be used. However, it is convenient to
consider an effective formation
rate~$\bar{\lambda}^F_{K_i}(\varepsilon)$ that leads to $dt$ fusion in
the muonic-molecular complex. The fusion probability depends on decay of
the created complex into the two initial objects: the muonic atom and
the hydrogen-isotope molecule. This process competes with transitions
leading to $dt$~fusion inside the complex. If the lifetime of the
complex ($\lesssim$~1~ns) is much shorter than its rotational relaxation
time, decay takes place back to the initial channel. When these times
are comparable, it is necessary to include back decay from lower
rotational states of the complex. In particular, in the limit of very
fast rotational relaxation, back decay from the ground rotational
state~$K_f=0$ is dominant. Such situation takes place in dense targets,
where interactions of the complex with neighboring molecules lead to
fast rotational deexcitation.  Calculations presented in
Refs.~\cite{ostr80,padi88} show that rotational relaxation of the
complex, via scattering on neighboring hydrogenic molecules, is fast at
the liquid hydrogen density. The effective formation rate is then
\begin{equation}
  \label{eq:form_eff}
  \bar{\lambda}^F_{K_i} = \sum_{K_f,S}
  \lambda^{SF}_{\nu_i K_i \nu_f K_f} \, 
  \mathcal{P}_{\text{fus}}^S \,, \qquad \nu_i=0 \,, 
\end{equation}
where $\mathcal{P}_{\text{fus}}^S$ is the fusion fraction 
\begin{equation*}
  \mathcal{P}_{\text{fus}}^S = 
  \lambda_f/\varGamma_S   
\end{equation*}
and the back-decay rate $\lambda^S_\text{bck}$ is given by
\begin{equation*}
  \lambda^S_\text{bck} = \sum_{F'} \varGamma_{SF'} \,, 
  \quad \varGamma_{SF'} = \sum_{\nu_i'}\sum_{K_i',K_f=0} 
  \varGamma^{SF'}_{\nu_fK_f,\nu_i'K_i'}   \,.
\end{equation*}
It is assumed here that the vibrational level~$\nu_f$ of the complex is
not changed during its lifetime. Though calculations of vibrational
relaxation of the muonic molecular systems in condensed targets have not
been performed yet, the available data~\cite{soue86} concerning
$\nu=1\to{}0$ relaxation time for H$_2$ in solid (8~$\mu$s) and liquid
(12~$\mu$s at 14.2~K) hydrogen suggest that such times are much greater
than the lifetime of the muonic molecular complex.

\subsection{Formation in a~solid in the strong-binding limit}
\label{sec:strong_bind}

Evaluation of the response function~$\mathcal{S}_\text{res}$ is
difficult, in a~general case. The first problem is that the operators
$\vec{R}_l(t)$, $\mathscr{H}$, and~$\varDelta\mathscr{H}$ in
Eq.~(\ref{eq:func_Y}) do not commute. However, when muonic molecule
formation takes place at energies significantly smaller than the mean
kinetic energy~$\mathscr{E}_T$ of molecule~D$C$, the perturbation
operator~(\ref{eq:deltaH}) is well approximated by its mean value
\begin{equation}
  \label{eq:res_shift}
  \begin{split}
    \varDelta\mathscr{H} \approx 
    \langle 0 \lvert \varDelta\mathscr{H} \rvert 0 \rangle 
    &=-\alpha \, \left\langle\boldsymbol{\nabla}^2_{R_l}
      /(2M_{\text{D}C})\right\rangle_T \\
    &= -\alpha\, \mathscr{E}_T  \equiv \varDelta\varepsilon_{if}< 0 \,.    
  \end{split}
\end{equation}
Using this approximation in~Eq.~(\ref{eq:func_Y}) we obtain 
\begin{equation*}
  \begin{split}
    \mathscr{Y}_\text{res}(\vec{k},t) 
    &\approx \exp(it\varDelta\varepsilon_{if})
    \bigl\langle\exp[-i\vec{k}\cdot\vec{R}_l(0)]  
    \exp[i\vec{k}\cdot\vec{R}_l(t)]\bigr\rangle_T \\
    &= \exp(it\varDelta\varepsilon_{if})\, \mathscr{Y}_{ll}(\vec{k},t).     
  \end{split}
\end{equation*}
Thus, function~$\mathscr{Y}_\text{res}$ reduces to the standard
incoherent correlation function
$\mathscr{Y}_{ll}(\vec{k},t)$~\cite{love84}, multiplied by the
factor~$\exp(it\varDelta\varepsilon_{if})$ describing a~variation of the
mean target energy due to its mass change. Hence, the formation
rate~(\ref{eq:resratsol_gen}) can be written down as follows:
\begin{equation}
  \label{eq:resratsol3}
  \begin{split}
    \lambda^{SF}_{\nu_iK_i,\nu_fK_f} = N_\text{mol} B_{if} & |V_{if}|^2 
    \int_{-\infty}^{\infty} dt\; \mathscr{Y}_{ll}(\vec{k},t) \\
    &\times \exp\bigl[-it(\varepsilon-\varepsilon_{if}) 
    -\tfrac{1}{2}\varGamma_S|t| \bigr] ,    
  \end{split}
\end{equation}
$\varepsilon_{if}$ being the resonance energy in the condensed target
\begin{equation}
  \label{eq:eres_sol}
  \varepsilon_{if}=\varepsilon^0_{if}+ \varDelta\varepsilon_{if} \,.
\end{equation}
This energy is shifted by $\varDelta\varepsilon_{if}<0$, compared to the
free-molecule resonance energy~$\varepsilon^0_{if}$. Note that such
a~resonant-energy shift was neglected in Refs.~\cite{lamb39,sing60},
where absorption of neutrons and $\gamma$-rays by heavy nuclei were
considered. An estimation of the shift in the case of $\gamma$~emission
from a~nucleus bound in a~solid, similar to Eq.~(\ref{eq:res_shift}) was
given in~Ref.~\cite{jose60}.

Using the following relation between $ \mathscr{Y}_{ll}(\vec{\kappa},t)$
and the standard single-particle function~$G_s(\vec{r},t)$~\cite{love84}:
\begin{equation*}
  G_s(\vec{r},t) = \frac{1}{(2\pi)^3} \int d^{\,3}\kappa \,
  \exp(-i\vec{\kappa}\cdot\vec{r})\, \frac{1}{N_\text{mol}}
  \sum_{l} \mathscr{Y}_{ll}(\vec{\kappa},t) \,,
\end{equation*}
the rate~(\ref{eq:resratsol3}) can be expressed as a~time and space Fourier 
transform 
\begin{equation}
  \label{eq:resratsol4}
  \begin{split}
    \lambda^{SF}_{\nu_iK_i,\nu_fK_f} = N_\text{mol}\, 
    B_{if}\, & |V_{if}|^2  \int  d^3 r\, dt\;  G_s(\vec{r},t) \\
    &\times \exp\left[i(\vec{\kappa}\cdot\vec{r}-\omega{}t) - 
      \tfrac{1}{2}\varGamma_S|t| \right] ,    
  \end{split}
\end{equation}
where the momentum transfer~$\vec{\kappa}$ and the energy
transfer~$\omega$ to the target are
\begin{equation}
  \label{eq:omega_def}
  \vec{\kappa} = \vec{k} \,, \qquad
  \omega=\varepsilon-\varepsilon_{if} \,.    
\end{equation}

Analogously, the back-decay width~(\ref{eq:elwidth_gen}) in the
strong-binding limit can be expressed by~$G_s(\vec{r},t)$ or by the
incoherent response function~$ \mathcal{S}_i$ introduced by Van Hove
\begin{equation*}
  \mathcal{S}_i(\vec{\kappa},\omega)=\frac{1}{2\pi}\int d^{\,3}r\,dt\;
  G_s(\vec{r},t)\,\exp\bigl[i(\vec{\kappa}\cdot
  \vec{r}-\omega t)\bigr]  \,. 
\end{equation*}
As a~result, Eq.~(\ref{eq:elwidth_gen}) takes a~simpler form
\begin{equation}
  \label{eq:elwidthsol1}
  \varGamma^{SF'}_{\nu_fK_f,\nu_iK_i} = 2\pi A_{if} 
  \int \frac{d^3k}{(2\pi)^3} \,\lvert V_{if}(\varepsilon)\rvert^2
  \, \widetilde{\mathcal{S}}_i(\vec{k},\omega') \,, 
\end{equation}
in which
\begin{equation}
  \label{eq:res_en_bck}
  \omega' = \tilde{\varepsilon}_{if}- \varepsilon \,,
  \qquad \tilde{\varepsilon}_{if} = 
  \varepsilon^0_{if} + \varDelta\tilde{\varepsilon}_{if} \,.
\end{equation}
and
\begin{equation}
  \label{eq:eres_cplx}
  \varDelta\tilde{\varepsilon}_{if} \equiv
  \langle \widetilde{n} \lvert \varDelta\mathscr{H}
  \rvert \widetilde{n} \rangle =  -\left(M_\text{cplx} / M_{\text{D}C}
    - 1\right)\, \widetilde{\mathscr{E}}_T < 0\,,
\end{equation}
$\widetilde{\mathscr{E}}_T$ being the mean kinetic energy
of the complex in the condensed target.

The equations derived above show that calculation of the formation and
back-decay rates in the low-energy limit reduces to evaluation of the
standard incoherent correlation functions, which are well-known in the
neutron scattering theory. In particular, for a~perfect gas or
a~harmonic solid composed of particles with mass~$M_{\text{mol}}$, these
functions take the simple Gaussian shapes~\cite{vanh54,love84}
\begin{equation}
  \label{eq:corr_gen}
   G_s(\vec{r},t) = \biggl[\frac{M_{\text{mol}}}{2\pi\gamma(t)}
   \biggr]^{3/2} \exp \biggl[
   -\frac{M_{\text{mol}}}{2\gamma(t)}\, r^2 \biggr] ,
\end{equation}
\begin{equation}
  \label{eq:Y_Gauss}
  \mathscr{Y}_{ll}(\vec{\kappa},t) = 
  \exp\left[-\gamma(t)\,\frac{\kappa^2}{2M_\text{mol}}\right] .
\end{equation}
For a~solid with a~cubic Bravais structure, function $\gamma(t)$ is 
\begin{equation}
  \label{eq:gamma_def}
  \begin{split}
    \gamma(t) =  \int_{0}^{\infty} dw \, \frac{Z(w)}{w}
    \bigl\{\coth(\tfrac{1}{2}\beta_T w) 
    \bigl[& 1-\cos(wt)\bigr] \\ 
    & -i\sin(wt)\bigr\} \,,    
  \end{split}
\end{equation}
where the normalized density of vibrational states $Z(w)$ has the
following properties:
\begin{equation}
  \label{eq:z_def}
  \begin{split}
    \int_{0}^{\infty} dw \, Z(w) &= 1 \,, \quad
    Z(w) = 0 \quad \text{ for } w > w_{\text{max}} \,, \\
    Z(-w) &\equiv Z(w)     
  \end{split}
\end{equation}
and $\beta_T=(k_\text{B} T)^{-1}$ ($k_\text{B}$ is Boltzmann's constant).

Solid hydrogens under low pressure, used for studies of muonic atoms and
molecules, are quantum molecular crystals. They have the Bravais fcc
polycrystalline structure or the hcp polycrystalline
structure~\cite{silv80,soue86}, for which Eqs.~(\ref{eq:corr_gen})
and~(\ref{eq:Y_Gauss}) are fair approximations.  As a~result, on
substitution $M_\text{mol}=M_{\text{D}C}$, we obtain the phonon expansion
for the resonant-formation rate
\begin{equation}
  \label{eq:resrat_gam_phon}
  \begin{split}
    \lambda^{SF}_{\nu_iK_i,\nu_fK_f} =&\,  N_\text{mol}\, 
    B_{if}\, |V_{if}|^2 \exp(-2W) \\
    &\times \left[
      \frac{\varGamma_S}{\omega^2+\tfrac{1}{4}\varGamma_S^2}
    +2\pi \sum_{n=1}^{\infty} g_{\varGamma{}n}(\omega) 
    \frac{(2W)^n}{n !} \right] ,    
  \end{split}
\end{equation}
in which
\begin{equation}
  \label{eq:phexp_gam}
  \begin{split}
    g_{\varGamma{}1}(w) &= \frac{1}{2\pi} \int_{-\infty}^{\infty} dz\,
    \frac{\varGamma_S}{z^2+\tfrac{1}{4}\varGamma_S^2} \,
    g_1(z+w,T) \,, \\
    g_{\varGamma{}n}(w) &= \int_{-\infty}^{\infty} dw'\,
    g_{\varGamma{}1}(w-w') \, g_{n-1}(w') \,, 
  \end{split}  
\end{equation}
and
\begin{equation}
  \label{eq:g_n}
  \begin{split}
    & g_1(w)  = \frac{1}{\gamma(\infty)} 
    \frac{Z(w)}{w} \left[\,n_{_\text{B}}(w)+1\right] \,, \\
    & g_n(w)  = \int_{-\infty}^{\infty} dw' \,
    g_1(w-w')\, g_{n-1}(w') \,, \\
    & \int_{-\infty}^{\infty} dw \, g_n(w) = 1 \,.
  \end{split}
\end{equation}
The exponent~$2W$ of the Debye-Waller factor $\exp(-2W)$, familiar in the
theory of neutron scattering, is 
\begin{equation*}
  \begin{split}
    2W(\kappa^2) & = \frac{\kappa^2}{2M_{\text{mol}}}\,\gamma(\infty) \\
    & = \frac{\kappa^2}{2M_{\text{mol}}} \int_0^{\infty} dw \,
    \frac{Z(w)}{w}\coth\left(\tfrac{1}{2} \beta_T w \right) \,,    
  \end{split}
\end{equation*}
where $\gamma(\infty)$ stands for the limit of~$\gamma(t)$
at~$t\to\infty$. Function $n_{_\text{B}}(w)$ denotes the Bose factor
\begin{equation}
  \label{eq:Bose_fac}   
  n_{_\text{B}}(w) = \left[\,\exp(\beta_T w)-1\right]^{-1}\,. 
\end{equation}

The Breit-Wigner term in expansion~(\ref{eq:resrat_gam_phon}) describes
recoil-less resonant formation. The sum with higher powers of~$2W$
correspond to quasiresonant muonic molecule formation with simultaneous
phonon creation or annihilation. In particular, the term with~$n=1$
describes formation connected with creation or annihilation of one
phonon. In the strong-binding limit $2W\ll{}1$, only few lowest terms
in~expansion~(\ref{eq:resrat_gam_phon}) are significant. The phonon
expansion~(\ref{eq:resrat_gam_phon}) is more general than an analogous
expansion in Ref.~\cite{sing60}, which includes the Breit-Wigner factor
only in the nonphonon term. This factor should be taken into account
also in the phonon terms, unless the natural resonance width is much
smaller than~$w_{\text{max}}$. For $2W\gtrsim{}1$, the
approximation~(\ref{eq:res_shift}) and Eq.~(\ref{eq:resrat_gam_phon})
are no longer valid.

When $dd\mu$ formation is concerned, the resonances are very narrow.
Thus, in this case, the limit $\varGamma_S\to{}0$ is practically
reached. The Breit-Wigner factor tends to the
$\delta$\nobreakdash-function profile and $g_{\varGamma{}n}\to{}g_n$. As
a~result, Eq.~(\ref{eq:resrat_gam_phon}) takes for $dd\mu$ a~simpler
form, derived in Ref.~\cite{adam01}
\begin{equation}
  \label{eq:resratsol5}
  \begin{split}
    \lambda^{SF}_{\nu_iK_i,\nu_fK_f} =& \,  2\pi  N B_{if} 
    \lvert V_{if} \rvert^2 \exp(-2W) \\
    &\times \left[\, \delta(\omega) +\sum_{n=1}^{\infty} g_n(\omega) 
      \frac{(2W)^n}{n !} \right] .     
  \end{split}
\end{equation}

A~phonon expansion can also be applied for estimation of the
back-decay rate. After integration of Eq.~(\ref{eq:elwidthsol1}) over
direction of~$\vec{k}$ one obtains
\begin{equation*}
  \varGamma^{SF'}_{\nu_fK_f,\nu_iK_i} = \frac{A_{if} }{\pi} 
  \int_{0}^{\infty} dk \, k^2 \, \lvert V_{if}(\varepsilon)\rvert^2
   \, \widetilde{\mathcal{S}}_i(k^2,\omega') \,.
\end{equation*}
Substitution of the phonon expansion for $\widetilde{\mathcal{S}}_i$
into equation above and then integration of the $\delta$-function term
lead to
\begin{equation}
  \label{eq:width_expan}
  \begin{split}
    \varGamma^{SF'}_{\nu_fK_f,\nu_iK_i} = \frac{A_{if} }{\pi}
    &\Bigg[ M_{a\mu}\, \widetilde{k}_{if} \lvert V_{if}
    (\tilde{\varepsilon}_{if})\rvert^2 \exp(-2\widetilde{W}_{if})\\
    &+\sum_{n=1}^{\infty} \int_{0}^{\infty} dk \, k^2 \,
    \lvert V_{if}(\varepsilon)\rvert^2 \exp(-2\widetilde{W}) \\ 
    & \qquad \qquad \times g_n(\omega')\, 
    \frac{(2\widetilde{W})^n}{n !} \Bigg] ,
  \end{split}
\end{equation}
in which
\begin{equation}
  \label{eq:2W_cplx}
  \begin{split}
    & 2\widetilde{W} = \frac{k^2}{2M_\text{cplx}} \, 
    \widetilde{\gamma}(\infty) \,, \quad
    2\widetilde{W}_{if} = 2\widetilde{W}(\widetilde{k}_{if}) \,, \\
    & \text{and}\quad 
    \widetilde{k}_{if} = \sqrt{\,2 M \tilde{\varepsilon}_{if}}     
  \end{split}
\end{equation}
are calculated for the harmonic lattice with the bound muonic molecular
complex. Note that Eq.~(\ref{eq:width_expan}) is valid only if a~main
contribution to the integral comes from small~$k$.

\subsection{Formation in the weak-binding limit}
\label{sec:form_weak_bind}

When the incident momentum of the muonic atom is large, the formation
time of a~muonic molecule is short compared to the characteristic time
scale of the dynamic response of the bulk target. Thus, a~contribution
to the response function~(\ref{eq:S_res}) from short times is dominant.
As a~result, it is sufficient to keep only linear terms in $t$ while
evaluating an asymptotic form of the correlation
function~$\mathscr{Y}_\text{res}(\vec{k},t)$. In calculations, we shall
use the following operator relation:
\begin{equation}
  \label{eq:exp_op}
  \exp(\hat{A})\exp(\hat{B}) = \exp(\hat{A}+\hat{B}+\hat{C}) \,, 
\end{equation}
where
\begin{equation*}
  \begin{split}
    \hat{C} = & \, \tfrac{1}{2}[\hat{A},\hat{B}] + 
    \tfrac{1}{12}\bigl[[\hat{A},\hat{B}],\hat{B}\bigr]+
    \tfrac{1}{12}\bigl[[\hat{B},\hat{A}],\hat{A}\bigr]\\  &+
    \tfrac{1}{24}\Bigl[\bigl[[\hat{B},\hat{A}],\hat{A}\bigr],\hat{B}\Bigr]
    +\ldots    
  \end{split}
\end{equation*}
Operator $\hat{C}=0$ only if $\hat{A}$ and $\hat{B}$ are commuting
operators.

The operators $\varDelta\mathscr{H}$ and $\mathscr{H}$, defined by
Eqs.~(\ref{eq:hamilt0}) and~(\ref{eq:hamiltn}), do not commute and the
operator~$\hat{C}$ in the expression
\begin{equation*}
  \exp\{it(\mathscr{H}+\varDelta\mathscr{H})\}\exp(-it\mathscr{H})=
  \exp(it\varDelta\mathscr{H}+\hat{C})
\end{equation*}
turns out to be a sum containing higher powers of~$t$. Since in this
approximation we restrict to terms linear with respect to~$t$ and to the
parameter~$\alpha\lesssim{}\tfrac{1}{2}$, the operator $\hat{C}$ in the
relation above can be neglected and thus the correlation function takes
the form
\begin{equation}
  \label{eq:Y_imp1}
  \begin{split}
    \mathscr{Y}_\text{res}(\vec{k},t) = \bigl\langle &
    \exp\{-i\vec{k}\cdot\vec{R}_l(0)\}
    \exp(it\varDelta\mathscr{H}) \\
    &\times \exp\{i\vec{k}\cdot\vec{R}_l(t)\} \bigr\rangle_T \,.    
  \end{split}
\end{equation}
Now we involve the basic approximation
\begin{equation}
  \label{eq:R_t_imp}
  \vec{R}_l(t) \approx \vec{R}(0) + (\vec{P}_l/M_{\text{D}C})\, t \,,
\end{equation}
where $\vec{P}_l$ denotes the momentum operator of the $l$th molecule.
This approximation is valid for $t\to{}0$. After substitution
of~Eq.~(\ref{eq:R_t_imp}) in Eq.~(\ref{eq:Y_imp1}) and multiple use of
the Eq.~(\ref{eq:exp_op}) we have
\begin{equation*}
  \begin{split}
    \mathscr{Y}_\text{res}(\vec{k},t) \approx & \,
    \exp\left(it\frac{k^2}{2M_\text{cplx}}\right)  
    \left\langle\exp\left(-it\alpha\frac{P_l^2}{2M_{\text{D}C}}\right)
    \right\rangle_T \\  &\times
    \left\langle\exp\left(it\frac{\vec{k}\cdot\vec{P}_l}{M_\text{cplx}}
      \right) \right\rangle_T \,,    
  \end{split}
\end{equation*}
Since the argument of the second exponential is small, we can use the
following approximation:
\begin{equation*}
  \begin{split}
    \left\langle\exp\left(-it\alpha\frac{P_l^2}{2M_{\text{D}C}}\right)
    \right\rangle_T & \approx \exp\left(-it\alpha\left\langle
        \frac{P_l^2}{2M_{\text{D}C}}\right\rangle_T \right) \\
    & = \exp(it\varDelta\varepsilon_{if}) \,    
  \end{split}
\end{equation*}
which involves the resonance-energy shift~(\ref{eq:res_shift}).
Substitution of the above equations in~Eq.~(\ref{eq:S_res}), with the
definitions~(\ref{eq:eres_sol}) and~(\ref{eq:omega_def}) taken into
account, leads to
\begin{equation}
  \label{eq:S_imp1}
  \begin{split}
    \mathcal{S}_\text{res}(\vec{\kappa},\omega) = \frac{1}{2\pi}
    \int_{-\infty}^{\infty} & dt\, \exp\left[-i\omega{}t - \tfrac{1}{2}
      \varGamma_S|t| +it\,\frac{\kappa^2}{2M_\text{cplx}}\right] \\
    &\times \left\langle\exp\left(it\,\frac{\vec{\kappa}\cdot\vec{P}_l}
        {M_\text{cplx}} \right) \right\rangle_T \,      
  \end{split}
\end{equation}
When the motion of the molecule~D$C$ is well described by an isotropic
harmonic potential, the Bloch identity 
\begin{equation}
  \label{eq:Bloch}
  \langle\exp{}\hat{Q}\rangle_T = 
  \exp\bigl(\tfrac{1}{2}\langle{}\hat{Q}^2\rangle_T\bigr) 
\end{equation}
may be applied for an operator~$\hat{Q}$ being a~linear combination of
the Bose operators of creation and annihilation. Since
momentum~$\vec{P}_l$ can be expressed by such operators (see
e.g.,~Ref~\cite{love84}), we have
\begin{equation}
  \label{eq:kp_harm}
  \begin{split}
    \left\langle\exp\left(it\,\frac{\vec{\kappa}\cdot\vec{P}_l}
        {M_\text{cplx}} \right) \right\rangle_T \, & = \,
    \exp(-\tfrac{1}{4}\varDelta_\text{res}^2) \,, \\
    \varDelta_\text{res}^2 & = \,\frac{2}{M_\text{cplx}^2}
    \bigl\langle(\vec{\kappa}\cdot\vec{P}_l)^2\bigr\rangle_T \,.    
  \end{split}
\end{equation}
In the case of cubic symmetry,
\begin{equation*}
  \bigl\langle(\vec{\kappa}\cdot\vec{P}_l)^2\bigr\rangle_T = 
  \tfrac{1}{3}\, \kappa^2 \langle{}P_l^2\rangle_T \,,
\end{equation*}
and this is a~fair approximation even for other lattices. Thus
\begin{equation*}
  \varDelta_\text{res}^2 = \frac{2}{3 M_\text{cplx}^2}\,  
  \kappa^2 \langle{}P_l^2\rangle_T =
  \frac{8}{3}\, \frac{M_{\text{D}C}}{M_\text{cplx}}\, 
  \left\langle\frac{P_l^2}{2M_{\text{D}C}}\right\rangle_{\! T} 
  \frac{\kappa^2}{2M_\text{cplx}} ,
\end{equation*}
which finally gives the following Doppler width:
\begin{equation}
  \label{eq:Dopp_res}
  \varDelta_\text{res} = 2\sqrt{\frac{2}{3}\, 
    \frac{M_{\text{D}C}}{M_\text{cplx}}\,
    \mathscr{E}_T\, \omega_R} \,,
\end{equation}
with the recoil energy
\begin{equation}
  \label{eq:recoil_cplx}
  \omega_\text{R} = \kappa^2/(2M_{\text{cplx}}) .
\end{equation}
In the case of a~solid hydrogen target, the mean kinetic energy of the
bound molecule equals
\begin{equation}
  \label{eq:meankin}
   \mathscr{E}_T = \tfrac{3}{2} \int_0^{\infty} dw \,
  Z(w)\,w \left[n_{_\text{B}}(w)+\tfrac{1}{2}\right] .
\end{equation}
This energy is much higher than $\mathscr{E}_T=\tfrac{3}{2}k_\text{B}T$
for a~corresponding Maxwellian gas, unless the temperature is
sufficiently high. This phenomenon was first taken into account by
Lamb~\cite{lamb39}, in resonant neutron absorption in solid crystals. In
particular, for a~low-pressure solid or liquid deuterium,
$\mathscr{E}_T\approx$~5~meV~\cite{momp96} due to a~large zero-point
motion of D$_2$ molecules in a~given target. The effective target
temperature~$T_\text{eff}$ corresponding to~$\mathscr{E}_T$ is then
defined as
\begin{equation}
  \label{eq:T_eff}
  T_{\text{eff}} \equiv \tfrac{2}{3} k_\text{B}^{-1} \mathscr{E}_T \,.
\end{equation}
For the considered solid D$_2$ case, $T_{\text{eff}}\approx{}40$~K.

Substitution of Eqs.~(\ref{eq:kp_harm}) and~(\ref{eq:recoil_cplx})
in Eq.~(\ref{eq:S_imp1}) leads to
\begin{equation}
  \label{eq:S_imp2}
  \begin{split}
    \mathcal{S}_\text{res}(\vec{\kappa},\omega) = \frac{1}{2\pi}
    \int_{-\infty}^{\infty} dt\,
    \exp \bigl[ & -i(\omega-\omega_\text{R})t \\
      & -\tfrac{1}{2}\varGamma_S|t| 
      -\tfrac{1}{4} \varDelta_\text{res}^2 t^2 \bigr]  \,.     
  \end{split}
\end{equation}
Then, applying the convolution theorem to the Fourier transform of
a~product, we obtain the asymptotic form of the resonance response
function
\begin{equation}
  \label{eq:S_imp3}
  \begin{split}
    \mathcal{S}_\text{res}(\vec{\kappa},\omega) = \frac{1}{2\pi^{3/2}}
    \, \frac{\varGamma_S}{\varDelta_\text{res}} &\int_{-\infty}^{\infty} 
    \frac{dz}{z^2 + \tfrac{1}{4}\varGamma_S^2} \\  &\times
    \exp\left[-\left(\frac{z+\omega-\omega_\text{R}}{\varDelta_\text{res}}
      \right)^{\!\!2}\right] .    
  \end{split}
\end{equation}
By virtue of Eq.~(\ref{eq:S_imp3}), the formation
rate~(\ref{eq:resratsol_gen}) in the weak-binding limit takes the form
\begin{equation}
  \label{eq:res_asym_gam}
  \begin{split}
    \lambda^{SF}_{\nu_iK_i,\nu_fK_f} = & \,
    N_\text{mol}\, B_{if}\, |V_{if}|^2 \,
    \frac{\varGamma_S}{\varDelta_\text{res}\sqrt{\pi}} \\
    &\times \int_{-\infty}^{\infty} 
    \frac{dz}{z^2 + \tfrac{1}{4}\varGamma_S^2} \, \exp
    \left[-\left(\frac{z+\omega-\omega_\text{R}}{\varDelta_\text{res}}
      \right)^{\!\!2}\right] .    
  \end{split}
\end{equation}
This equation is similar (apart from the muonic-molecule
factor~$N_\text{mol}B_{if}|V_{if}|^2$) to the expression for resonant
absorption of neutrons in a~gas target, obtained by Bethe and
Placzek~\cite{beth37}. However, the resonance width~(\ref{eq:Dopp_res})
and recoil energy~(\ref{eq:recoil_cplx}) take into account a~change of
the target particle mass in the absorption process, which is neglected
in their work.

In the limit $\varGamma_S\to{}0$, Eqs.~(\ref{eq:S_imp3})
and~(\ref{eq:res_asym_gam}) tend to the following expressions:
\begin{equation}
  \label{eq:S_res_gauss}
  {\mathcal S}_\text{res}(\vec{\kappa},\omega) = 
  \frac{1}{\varDelta_\text{res} \sqrt{\,\pi}} \,
  \exp \left[-\left(\frac{\omega - \omega_\text{R}}
    {\varDelta_\text{res}} \right)^{\!2} \,\right]
\end{equation}
and
\begin{equation}
  \label{eq:res_asym_gauss}
  \begin{split}
    \lambda^{SF}_{\nu_iK_i,\nu_fK_f} = & \, 2\sqrt{\pi}\,
    N_\text{mol}\, B_{if}\, |V_{if}|^2 \\
    &\times \frac{1}{\varDelta_\text{res}} 
    \exp\left[-\left(
        \frac{\omega-\omega_\text{R}}{\varDelta_\text{res}}
      \right)^{\!\!2}\right] ,    
  \end{split}
\end{equation}
respectively. Function~(\ref{eq:S_res_gauss}) has the Gaussian form,
identical with that used for description of incoherent scattering at
large energies. However, the Doppler width~(\ref{eq:Dopp_res}) and the
recoil energy~(\ref{eq:recoil_cplx}) in~$\mathcal{S}_\text{res}$ are
different from the corresponding variables
\begin{equation}
  \label{eq:Dopp_width}
  \varDelta_\text{R} = 
  2\sqrt{\tfrac{2}{3}\, \mathscr{E}_T \, \omega_\text{R}} 
\end{equation}
and 
\begin{equation}
  \label{eq:recoil_mol}
  \omega_\text{R} = \kappa^2/(2M_{\text{mol}}) \,,
\end{equation}
which determine the asymptotic form of the standard incoherent response
function~$\mathcal{S}_i$~\cite{love84}. Function
$\mathcal{S}_\text{res}$ tends to~$\mathcal{S}_i$ if the approximation
$M_\text{cplx}\approx{}M_{\text{D}C}$ is valid. However, in the case of
muonic molecule formation, this is only a~rough approach because the
mass of a~muonic hydrogen atom is comparable with the mass of a~hydrogen
isotope molecule. Note that, in the strong-binding limit, mass
$M_\text{cplx}$ enters only the resonance-energy
shift~(\ref{eq:res_shift}). The phonon
expansion~(\ref{eq:resrat_gam_phon}) is expressed in terms of
$W\sim{}\kappa^2/2M_{\text{D}C}$, not in terms of
$\widetilde{W}\sim{}\kappa^2/2M_\text{cplx}$. The reason is that
Eq.~(\ref{eq:resrat_gam_phon}) is valid for small collision energies,
when the target molecule is strongly bound in the lattice. Therefore,
the momentum is mostly transfered to the whole crystal.

Function~(\ref{eq:S_res_gauss}) can be used for evaluation of the
back-decay rate, if large final momenta give main contribution to the
integral~(\ref{eq:elwidth_gen}). After integration over direction
of~$\vec{k}$ in Eq.~(\ref{eq:elwidth_gen}), with the asymptotic
function~(\ref{eq:S_res_gauss}) inserted, one obtains
\begin{equation}
  \label{eq:width_gauss}
  \begin{split}
    \varGamma^{SF'}_{\nu_fK_f,\nu_iK_i} = 
    \frac{A_{if}}{\pi^{3/2}\widetilde{\varDelta}_\text{res}} &
    \int_{0}^{\infty} dk\, k^2 \,\lvert V_{if}(\varepsilon)\rvert^2 \\
    &\times \exp\left[-\left(\frac{\omega'-\omega'_R} 
        {\widetilde{\varDelta}_\text{res}}\right)^2\right] ,    
  \end{split}
\end{equation}
where $\omega'$ is defined by Eq.~(\ref{eq:res_en_bck}). The parameters
$\widetilde{\varDelta}_\text{res}$ and~$\omega'_R$ are calculated from
Eqs.~(\ref{eq:Dopp_res}) and~(\ref{eq:recoil_cplx}), using the
replacements $M_{\text{D}C}\leftrightarrow{}M_\text{cplx}$ and
$\mathscr{E}_T\to\widetilde{\mathscr{E}}_T$.

Let us note that Eqs.~(\ref{eq:S_imp3}) and~(\ref{eq:S_res_gauss}) are
general since they are derived in the impulse
approximation~(\ref{eq:R_t_imp}) without using specific properties of
a~given target, apart from the single parameter~$\mathscr{E}_T$.
Therefore, they are valid for liquid and dense gaseous hydrogens. They
can be also used for description of resonant absorption processes other
than muonic molecule formation, when a~mass change cannot be neglected.
In such a~case, $\varGamma_S$ should be replaced by an appropriate
natural resonance width.

\subsection{Formation in a~solid at intermediate energies}
\label{sec:form_interm}

The formation rate calculated according the asymptotic
form~(\ref{eq:res_asym_gam}) becomes very inaccurate when the collision
energy is much greater than the maximal frequency~$w_\text{max}$ of
a~crystal. In particular, this concerns recoil-less formation, which is
dominant at lowest energies. Therefore, at the intermediate energies, it
is reasonable to represent the formation rate as a~sum of the exact
nonphonon term from expansion~(\ref{eq:resrat_gam_phon}) and the
subsequent phonon terms, which we obtain below in the impulse
approximation. Using Eqs.~(\ref{eq:exp_op}) and~(\ref{eq:Bloch}), it can
be shown that the relation
\begin{equation}
  \label{eq:Y_res_class}
  \begin{split}
    \mathscr{Y}_{ll}^\text{res}(\vec{k},t) \approx & \,
    \mathscr{Y}_{ll}(\vec{k},t) \exp(it\varDelta\varepsilon_{if}) 
    \\ &\times
    \exp\biggl\{\alpha\left[it -\tfrac{2}{3}(\alpha+2)\,\mathscr{E}_T 
      \, t^2\right]\frac{k^2}{2M_{\text{D}C}}\biggr\}     
  \end{split}
\end{equation}
is valid in this approximation. Inserting Eqs.~($\ref{eq:Y_Gauss}$)
and (\ref{eq:Y_res_class}) into~Eq.~(\ref{eq:S_res}) we have
\begin{equation}
  \label{eq:S_res_class}
  \begin{split}
    \mathcal{S}_\text{res}(\vec{\kappa},\omega) = 
    \frac{1}{2\pi} \int_{-\infty}^{\infty} & dt\,  
    \exp\biggl\{ -it\omega-\tfrac{1}{2}\varGamma_S|t| \\ 
    +\bigl[&-\gamma(t) + i\alpha{}t \\ &-\tfrac{2}{3}\alpha
      (\alpha+2)\,\mathscr{E}_T\, t^2\bigr]
    \frac{\kappa^2}{2M_{\text{D}C}}\biggr\} .    
  \end{split}
\end{equation}
Substituting the approximation
$\gamma(t)\approx{}-it+\tfrac{2}{3}\,\mathscr{E}_T{}t^2$ for short times
into~Eq.~(\ref{eq:S_res_class}) and integrating over~$t$ yields the
asymptotic form~(\ref{eq:S_imp3}) of the response function. However, 
we now expand Eq.~(\ref{eq:S_res_class}) in powers of~$\kappa^2$
\begin{equation}
  \label{eq:S_inter1}
  \begin{split}
    \mathcal{S}_\text{res}(\vec{\kappa},\omega) =& \, 
    \frac{1}{2\pi}\exp(-2W)\sum_{n=0}^{\infty}
    \frac{(2W)^n}{n!} \\ &\times
    \int_{-\infty}^{\infty} dt\, \exp(-i\omega{}t
    -\tfrac{1}{2}\varGamma_S|t|)\,[\mathscr{F}(t)]^n , \\
    \mathscr{F}(t) &=
    1+i\,\frac{1+\alpha}{\gamma(\infty)}\, t  -
    \frac{2}{3}\, \frac{(1+\alpha)^2}{\gamma(\infty)}\, 
    \mathscr{E}_T\, t^2 . 
  \end{split}
\end{equation}
Function $g_n$ is defined by Eq.~(\ref{eq:g_n}). The integral
over~$t$ is estimated using the exponential approximation to
function~$\mathscr{F}$
\begin{equation}
  \label{eq:aprox_gauss}
  \mathscr{F}(t) \approx \exp(x) \,, \qquad  x \approx
  \frac{it}{\gamma_\alpha} - \tfrac{1}{2}\,\varDelta_\alpha^2\,t^2 \,,
\end{equation}
where~$x$ contains only leading terms in~$t$ and
\begin{equation*}
  \gamma_\alpha\equiv\frac{M_\text{cplx}}{M_{\text{D}C}}\,\gamma(\infty)
  \,, \qquad
  \varDelta_\alpha^2 \equiv \frac{4}{3}\,\frac{M_{\text{D}C}}%
  {M_\text{cplx}}\, \frac{\mathscr{E}_T}{\gamma_\alpha} 
  - \frac{1}{\gamma_\alpha^2} \,.
\end{equation*}
Then, integration in Eq.~(\ref{eq:S_inter1}), using the convolution
theorem, leads to
\begin{equation}
  \label{eq:S_inter2}
  \begin{split}
    \mathcal{S}_\text{res}(\vec{\kappa},\omega) =  
    \exp(-2W)\Biggl[ & \frac{1}{2\pi} \, \frac{\varGamma_S}
    {\omega^2 +\tfrac{1}{4}\varGamma_S^2} \\
    & +\sum_{n=1}^{\infty}
    \mathfrak{g}_n(\omega)\,\frac{(2W)^n}{n!} \Biggr] ,     
  \end{split}
\end{equation}
where
\begin{equation*}
  \begin{split}
    \mathfrak{g}_1(\omega) = \frac{1}{2\pi}\int_{-\infty}^{\infty} & dz\,
    \frac{\varGamma_S}{z^2+\tfrac{1}{4}\varGamma_S^2} \, 
    \frac{Z(z+\omega_\alpha)}{z+\omega_\alpha} \\
    &\times \left[n_{_\text{B}}(z+\omega_\alpha)+1\right], \qquad
    \omega_\alpha = \frac{\omega}{1+\alpha} \,,     
  \end{split}
\end{equation*}
and, for $ n\geq{}2$,
\begin{equation*}
  \begin{split}
    \mathfrak{g}_n(\omega) = \frac{1}{(2\pi)^{3/2}} \,
    \frac{\varGamma_S}{n^{1/2}\varDelta_\alpha}
    \int_{-\infty}^{\infty} & dz\,
    \frac{1}{z^2+\tfrac{1}{4}\varGamma_S^2} \\
    &\times \exp
    \left[\frac{(z+\omega-n/\gamma_\alpha)^2}{2n\varDelta_\alpha^2}\right].
  \end{split}
\end{equation*}
The first term of Eq.~(\ref{eq:S_inter1}) has been replaced
in~(\ref{eq:S_inter2}) by the exact Breit-Wigner term.  Also the
one-phonon ($n=1$) contribution to~$\mathcal{S}_\text{res}$, is replaced
here by a~more accurate term depending on~$\mathfrak{g}_1$.
Function~$\mathfrak{g}_1$ is calculated on substitution of the exact
function~$\gamma(t)$ for a~harmonic solid
into~Eq.~(\ref{eq:S_res_class}). Every multiphonon term
in~Eq.~(\ref{eq:S_inter2}) is represented by the convolution of the
Breit-Wigner profile with a~Gaussian obtained using
Eq.~(\ref{eq:aprox_gauss}). It now follows that
\begin{equation}
  \label{eq:resrat_inter}
  \begin{split}
    \lambda^{SF}_{\nu_iK_i,\nu_fK_f} = & \,
    N_\text{mol}\, B_{if}\, |V_{if}|^2 \, \exp(-2W) \\
    &\times \left[\frac{\varGamma_S}
      {\omega^2 + \tfrac{1}{4}\varGamma_S^2} + 2\pi \sum_{n=1}^{\infty}
      \mathfrak{g}_n(\omega)\, \frac{(2W)^n}{n!} \right] .       
  \end{split}
\end{equation}
The form of this expansion is similar to that
of~Eq.~(\ref{eq:resrat_gam_phon}), derived in the strong-binding limit.
However, functions~$\mathfrak{g}_n$ are obtained in the impulse
approximation and they are different from the corresponding functions
$g_{\varGamma{}n}$ given by Eq.~(\ref{eq:phexp_gam}).
For the one-phonon term, we have %
$\mathfrak{g}_1(\omega)=g_{\varGamma{}1}(\omega_\alpha)$, which is the
direct result of using the exact $\gamma(t)$ in derivation
of~$\mathfrak{g}_1$. Thus, Eqs.~(\ref{eq:resrat_inter})
and~(\ref{eq:resrat_gam_phon}) give the same rate at smallest energy
transfers. At large~$\varepsilon$, when many multiphonon terms are
important, the target response no longer displays a~rich structure.  The
rate~(\ref{eq:resrat_inter}) tends therefore to the simpler
form~(\ref{eq:res_asym_gam}), which is characterized by the recoil
energy~(\ref{eq:recoil_cplx}) with the correct mass~$M_\text{cplx}$.

In the limit $\varGamma_S\to\nobreak{}0$, the
rate~(\ref{eq:resrat_inter}) takes the form similar to
Eq.~(\ref{eq:resratsol5})
\begin{equation}
  \label{eq:resrat_inter_delta}
  \begin{split}
    \lambda^{SF}_{\nu_iK_i,\nu_fK_f} = & \, 2\pi  N B_{if} 
    |V_{if}|^2 \exp(-2W) 
    \\  &\times
    \Biggl[ \delta(\omega)+\sum_{n=1}^{\infty} \mathfrak{g}_n(\omega,T)\, 
    \frac{(2W)^n}{n !} \Biggr] \,,     
  \end{split}
\end{equation}
with the expansion coefficients
\begin{equation*}
  \begin{split}
    \mathfrak{g}_1(\omega) &= \frac{Z(\omega_\alpha)}{\omega_\alpha}\,
    \left[n_{_\text{B}}(\omega_\alpha)+1\right] \,, \\ 
    \mathfrak{g}_n(\omega) &=
    \frac{1}{(2\pi n)^{1/2}\varDelta_\alpha}\,  \exp
    \left[\frac{(\omega-n/\gamma_\alpha)^2}{2n\varDelta_\alpha^2}\right],
    \qquad n\geq 2 \,.
  \end{split}
\end{equation*}

The back-decay rate can be calculated analogously. The result is given
by Eq.~(\ref{eq:width_expan}) with functions $g_n$ replaced by the
corresponding functions~$\mathfrak{g}_n$ from
~Eq.~(\ref{eq:resrat_inter_delta}). It is also necessary to make
the following substitutions: %
$M_{\text{D}C}\leftrightarrow{}M_\text{cplx}$ and
$\mathscr{E}_T\to\widetilde{\mathscr{E}}_T$.

\section{Results of calculations for $\boldsymbol{dt\mu}$ 
formation in solid hydrogens}
\label{sec:dtm-sol}

In this Section, the rates of resonant $dt\mu$ formation in solid HD,
D$_2$, and DT are calculated. It is assumed that these targets are kept
at zero or low pressures ($\ll{}10$~kbar), which corresponds to the
TRIUMF or RIKEN-RAL experimental conditions. Measurements of the
formation rates at TRIUMF have been performed using energetic
($\sim{}1$~eV) beams of $t\mu$ atoms. Therefore, the rates are evaluated
here in a~wide energy interval $\varepsilon\lesssim{}1$~eV. This
involves resonant $dt\mu$ formation with simultaneous excitations of
a~few lowest vibrational levels of the muonic-molecular complex. The
values of the rates are given for a~normalized target density of
$4.25\times{}10^{22}$~atoms/cm$^3$ (liquid-hydrogen density).

Since exact forms of the vibrational-state distribution~$Z(w)$ for the
experimental polycrystalline targets are not known, the Debye model of
an isotropic solid has been used in the calculations presented below.
The values of the Debye temperature~$\varTheta_\text{D}$ are taken from
the available literature~\cite{silv80,soue86}.

The resonance energies and energy-dependent transition-matrix elements
for isolated target molecules HD, D$_2$, and~DT, calculated according to
the method presented in Ref.~\cite{faif96}, are the starting point for
evaluation of the formation rates in solid hydrogens. The
transition-matrix elements are available for the rotational transitions
$K_i=0, 1\to{}K_f=0,\ldots, 9$.

Resonant $dt\mu$ formation in a~bound D$_2$ molecule is the most
complicated case. The lowest resonances, corresponding to the
vibrational transition $\nu_i=0\to\nu_f=2$ and different rotational
states $K_i$ and~$K_f$, are located in the vicinity of $\varepsilon=0$
with the radius of a~few tens~meV. The resonance energies in this
region, for a~free D$_2$ molecule and for a~D$_2$ bound in a~3-K solid
deuterium, are shown in~Table~\ref{tab:eres_dtm}.
\begin{table}[htb]
  \begin{center}
    \caption{Resonance energies for $dt\mu$ formation in 
      $t\mu$ scattering from a~free D$_2$
      molecule~($\varepsilon_{if}^0$) and from a~3-K solid-D$_2$ target
      ($\varepsilon_{if}$), corresponding to the vibrational transition
      $\nu_i=0\to{}\nu_f=2$. These energies are given in the
      corresponding center-of-mass systems.
      \label{tab:eres_dtm}}
    \begin{ruledtabular}
    \newcolumntype{.}{D{.}{.}{2.4}}
    \begin{tabular}{. . c c c c}
      \multicolumn{1}{c}{$\varepsilon_{if}^0$ (meV)}&
      \multicolumn{1}{c}{$\varepsilon_{if}$ (meV)}& 
      \multicolumn{1}{c}{$F$}&    
      \multicolumn{1}{c}{$K_i$}&
      \multicolumn{1}{c}{$K_f$}&
      \multicolumn{1}{c}{$S$}    \\
      \hline
      -25.66  & -27.95  &  1  & 1 & 4 &  1  \\
      -21.25  & -23.54  &  1  & 0 & 4 &  0  \\
      -18.66  & -20.95  &  1  & 1 & 4 &  2  \\
      -18.25  & -20.54  &  1  & 0 & 4 &  1  \\
      -11.25  & -13.54  &  1  & 0 & 4 &  2  \\
      \hline
      -24.15  & -26.44  &  0  & 1 & 0 &  1  \\
      -19.28  & -21.57  &  0  & 1 & 1 &  1  \\
      -16.74  & -19.02  &  0  & 0 & 0 &  1  \\
      -11.86  & -14.15  &  0  & 0 & 1 &  1  \\
      -9.547  & -11.84  &  0  & 1 & 2 &  1  \\
      -2.133  & -4.423  &  0  & 0 & 2 &  1  \\
       5.007  &  2.718  &  0  & 1 & 3 &  1  \\
       12.42  &  10.13  &  0  & 0 & 3 &  1  \\
       24.34  &  22.05  &  0  & 1 & 4 &  1  \\
       31.75  &  29.46  &  0  & 0 & 4 &  1  \\
     \end{tabular}
   \end{ruledtabular}
 \end{center}
\end{table}
In particular, there are several subthreshold resonances that give
significant contributions to the low-energy rates, because of wide
resonance profiles. The resonance-energy shift~(\ref{eq:res_shift}) for
a~deuterium target at~3~K is $\varDelta\varepsilon_{if}=-2.29$~meV.
Resonances in the upper spin state~$F=1$ have much smaller energies than
those for~$F=0$ with the same rotational quantum numbers. In particular,
the largest values of~$\varepsilon_{if}$ for $F=1$, shown
in~Table~\ref{tab:eres_dtm}, are due to the excitations
$K_i=0,~1\to{}K_f=4$. The only matrix elements that do not tend to
zero at $\varepsilon\to{}0$ correspond to the dipole transitions
$K_i=0\to{}K_f=1$ and $K_i=1\to{}K_f=0$,~2. For $F=1$, all these
transitions are associated with $\varepsilon_{if}<-50$~meV and thus they
give very small contribution to the resonant $dt\mu$-formation rate. As
a~result, the low-energy rate is determined mainly by $t\mu$ scattering
in the $F=0$ state. However, even for $F=0$, the dipole transitions are
connected with negative resonance energies, though much closer to
$\varepsilon=0$ than in the $F=1$ case.  The lowest positive resonances
appear in the transitions $K_i=0\to{}K_f=3$,~4 and $K_i=1\to{}K_f=3$,~4.
They are characterized by strongly varying transition-matrix
elements~\cite{faif96}, which is illustrated in~Figs.~\ref{fig:v0f_dtm}
and~\ref{fig:v1f_dtm}. 
\begin{figure}[htb]
  \begin{center}
    \includegraphics[width=7cm]{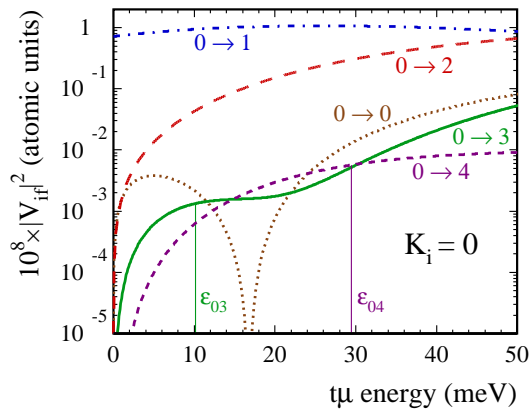} 
    \caption{Matrix elements~$|V_{if}(\varepsilon)|^2$ 
      versus $t\mu$ energy for the transitions
      $K_i=$~0~$\to{}K_f=$~0,~1,~2,~3,~4 and $\nu_i=0\to{}\nu_f=2$. The
      vertical lines denote energies~$\varepsilon_{if}$ of the lowest
      resonances. Labels ``$i\to{}f$'' stand for the rotational
      transitions $K_i\to{}K_f$.
    \label{fig:v0f_dtm}}
  \end{center}
\end{figure}
\begin{figure}[htb]
  \begin{center}
    \includegraphics[width=7cm]{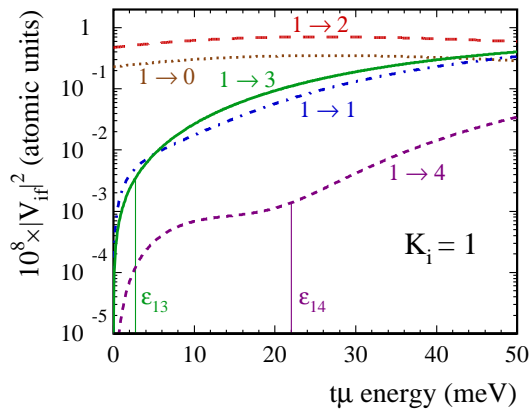}
    \caption{Matrix elements~$|V_{if}(\varepsilon)|^2$ 
      versus $t\mu$ energy for the
      transitions$K_i=$~1~$\to{}K_f=$~0,~1,~2,~3,~4 and
      $\nu_i=0\to{}\nu_f=2$. Notation is the same as in
      Fig.~\ref{fig:v0f_dtm}.
    \label{fig:v1f_dtm}}
  \end{center}
\end{figure}
Let us note that this situation is very different from the $dd\mu$ case,
where low-energy formation is determined by the dipole transitions, with
the matrix elements slowly varying below a~few
tens~meV~\cite{faif96,adam01}. Another difference between the $dd\mu$
and $dt\mu$ case is involved by larger separations of the neighboring
$dt\mu$ resonances corresponding to $K_i=0$ and~$K_i=1$. Therefore, for
$dt\mu$ one can expect more pronounced differences between resonant
formation in solid ortho-D$_2$ and para-D$_2$ than those found for
$dd\mu$ case~\cite{toyo03}. Most pure-deuterium experiments in $\mu$CF
have been carried out in targets with the statistical mixture of ortho
and parastates (called ``normal'' deuterium nD$_2$, according to the
nomenclature used in~Ref.~\cite{soue86}).

\begin{figure}[htb]
  \begin{center}
    \includegraphics[width=7cm]{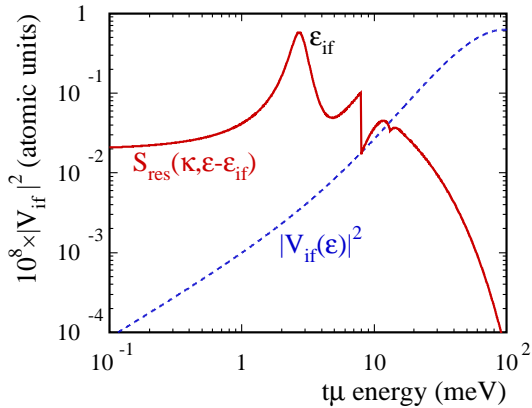}
    \caption{Transition-matrix element~$|V_{if}(\varepsilon)|^2$ for 
      resonant $dt\mu$ formation (transition $\nu_i=0\to{}\nu_f=2$,
      $K_i=1\to{}K_f=3$, dashed line) and the response function
      $\mathcal{S}_\text{res}(\vec{\kappa},\varepsilon-\varepsilon_{if})$
      (in arbitrary units, solid line) for the resonance $F=0\to{}S=1$
      in 3-K para-D$_2$. The peak of the Breit-Wigner term from
      Eq.~(\ref{eq:S_inter2}) is centered at the resonance energy
      $\boldsymbol{\varepsilon_{if}}=$~2.7~meV.
    \label{fig:respv2_dtm}}
  \end{center}
\end{figure}
In~Fig.~\ref{fig:respv2_dtm} is shown~$|V_{if}(\varepsilon)|^2$ for the
transition $\nu_i=0\to{}\nu_f=2$, $K_i=1\to{}K_f=3$, together with the
response function~(\ref{eq:S_inter2}) for the resonance $F=0\to{}S=1$
located at~$\varepsilon_{if}=2.7$~meV. The phonon terms
in~$\mathcal{S}_\text{res}$ are calculated assuming $\varGamma_S=0$,
since in this example we want to neglect their convolution with the
Breit-Wigner profile. There is a~strong contrast between resonant
formation of the molecules $dt\mu$ and $dd\mu$~\cite{adam01} in a~solid
deuterium. In the $dt\mu$ case, the wide Breit-Wigner peak is not so
much pronounced as the narrow recoil-less $dd\mu$ resonances. The matrix
element~$|V_{if}(\varepsilon)|^2$ raises by a~few orders of magnitude
within the width of 100~meV of the multiphonon distribution. Thus, the
phonon contribution to the $dt\mu$-formation rate is comparable with the
nonphonon one, already above a~few~meV. This means that a~detailed form
of the density~$Z(w)$ of vibrational lattice states is necessary for
accurate calculation of the low-energy $dt\mu$-formation rate in
a~solid~D$_2$.
\begin{figure}[htb] 
  \begin{center}
    \includegraphics[width=7cm]{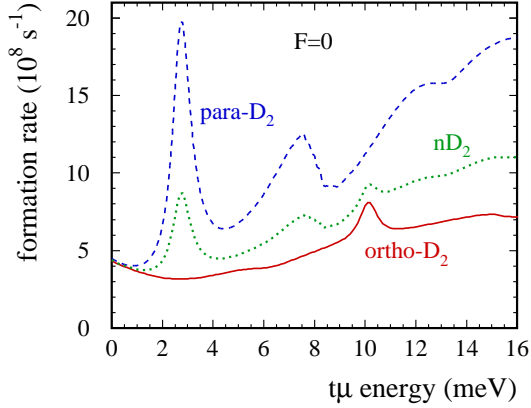}
    \caption{Low-energy $dt\mu$-formation rate for~$F=0$ in
      a~3-K solid nD$_2$ (``normal'' deuterium~\cite{soue86}),
      ortho-D$_2$, and~para-D$_2$, calculated using
      Eq.~(\ref{eq:resrat_inter}).
    \label{fig:dtm_d2_sh3kf0}}
  \end{center}
\end{figure}
\begin{figure}[htb] 
  \begin{center}
    \includegraphics[width=7cm]{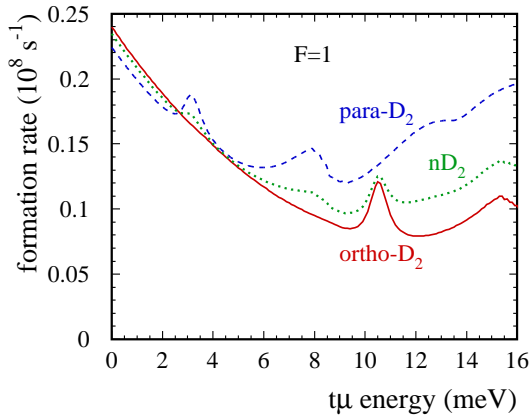}
    \caption{Low-energy $dt\mu$-formation rate for~$F=1$ in
      the same targets as in Fig.~\ref{fig:dtm_d2_sh3kf0}.
    \label{fig:dtm_d2_sh3kf1}}
  \end{center}
\end{figure}
A~shape of the phonon spectrum in the energy-dependent rate is strongly
distorted, which one sees in~Fig.~\ref{fig:dtm_d2_sh3kf0} evaluated
using Eq.~(\ref{eq:resrat_inter}). Nevertheless, the one-phonon and
two-phonon terms are clearly distinguished in the curve corresponding to
para-D$_2$. In ortho-D$_2$, the resonance with the lowest
$\varepsilon_{if}>0$ is located at 10~meV.  Therefore, the Breit-Wigner
peak is strongly suppressed by the Debye-Waller factor and the rate is
quite flat. At $\varepsilon\to{}0$, the rates are determined by the
wings of the Breit-Wigner peaks, because phonon contribution to the
rates vanishes when $\kappa$~approaches zero.  For $F=1$, the main
resonances are far from the considered low-energy interval (see
Table~\ref{tab:eres_dtm}). The rates shown
in~Fig.~\ref{fig:dtm_d2_sh3kf1} are thus determined by the Breit-Wigner
wings of the deep subthreshold resonances with small contributions from
the weak resonance ($K_f=5$) located at $\varepsilon_{if}>0$. As
a~result, the formation rates for~$F=1$ are lower by two orders of
magnitude than those for~$F=0$.

\begin{figure}[htb] 
  \begin{center}
    \includegraphics[width=7cm]{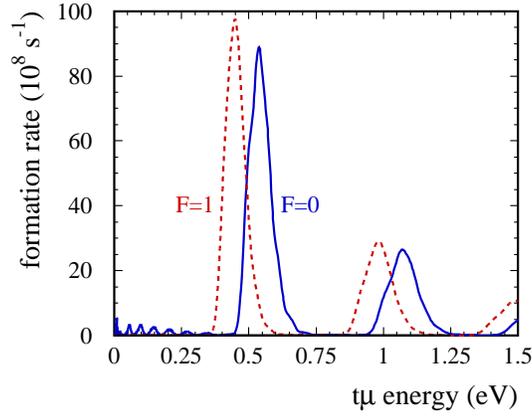}
    \caption{Resonant $dt\mu$-formation rate in 3-K gaseous~nD$_2$ 
      calculated in the laboratory frame.
    \label{fig:dtm_d2gas3k}}
  \end{center}
\end{figure}

Resonances in $t\mu$ scattering from~D$_2$, corresponding to the
vibrational excitations $\nu_f\geq{}3$ of the [$(dt\mu)dee$] complex,
are located at higher energies $\varepsilon\gtrsim{}0.2$~eV. Therefore,
they are well described by the asymptotic form~(\ref{eq:res_asym_gam}),
which is independent of~$Z(w)$. Therefore, the formation rate is
determined accurately using only the mean kinetic energy~$\mathscr{E}_T$
of a~D$_2$ molecule.
\begin{figure}[htbp] 
  \begin{center}
    \includegraphics[width=7cm]{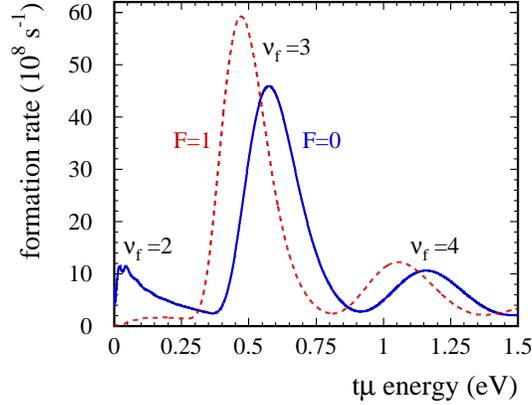}
    \caption{Resonant $dt\mu$-formation rate in a~3-K solid nD$_2$
      for $F=0$ and~$F=1$. The label~$\nu_f$ denotes the vibrational
      state of the created $[(dt\mu)dee]$ complex.
      \label{fig:dtm_nd2_3k}}
  \end{center}
\end{figure}
The formation rate in a~3-K solid nD$_2$ is plotted
in~Fig.~\ref{fig:dtm_nd2_3k}, for several~$\nu_f$. For comparison, in
Fig.~\ref{fig:dtm_d2gas3k} is shown the $dt\mu$-formation rate for 3-K
gaseous~nD$_2$. The energy-dependent rate for a~perfect deuterium gas
has been calculated assuming a~3-K Maxwellian distribution of the D$_2$
kinetic energy. This rate includes only formation due to two-body
$t\mu+$D$_2$ collisions.
The resonant-formation rates presented in~Figs.~\ref{fig:dtm_nd2_3k}
and~\ref{fig:dtm_d2gas3k} display a~striking difference between the gas
and the solid case. At $\varepsilon\to{}0$, the theory developed for
two-body collisions in a~perfect gas gives a~negligible resonant
formation rate. This result disagrees with the average formation rates
determined by measurements performed in liquid and cold dense-gas
targets~\cite{breu87} and in solid~\cite{kawa03} targets. The rate for
the solid shows a~strong contribution from the subthreshold resonances,
which leads to a~large rate in the limit $\varepsilon\to{}0$.
Solid-state effects are also significant at higher energies. The
resonance peaks in the solid are much broader than those in the gas
because of a~large effective target temperature connected with strong
zero-point motion of D$_2$ in solid deuterium~\cite{momp96}. The widths
of the peaks increase with rising recoil energy. However, the centers of
higher-energy peaks in the both targets have similar locations since, in
the impulse-approximation limit, the recoil
energy~(\ref{eq:recoil_cplx}) for the muonic-molecular complex bound in
a~solid equals to that for the isolated complex. A~small
difference~$\varDelta\varepsilon_{if}$ of the resonance energy between
the solid and the gas is negligible for $\varepsilon_{if}\gg{}1$~meV.

\begin{figure}[htb] 
  \begin{center}
    \includegraphics[width=7cm]{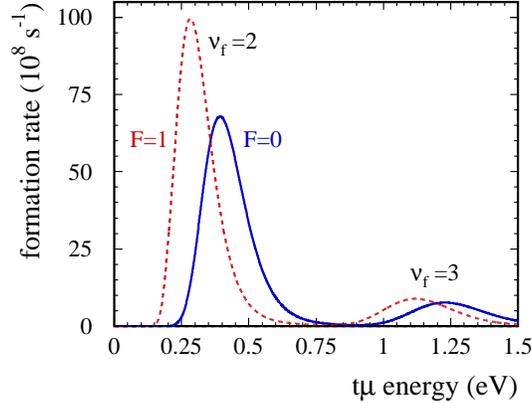}
    \caption{Resonant $dt\mu$-formation rate in 3-K solid~HD. 
      \label{fig:dtm_hd3k}}
  \end{center}
\end{figure}

Calculation of the $dt\mu$-formation rate for a~solid HD or DT is
simpler than for~D$_2$ since in the HD or DT case there are no
significant resonances in the close vicinity of $\varepsilon=0$. This is
caused by different values of the rotational and vibrational quanta for
these three molecules. The HD molecule is the lightest one and the
resonances connected with~$\nu_f=2$ are situated in~HD
above~0.1~eV~\cite{faif91,faif96}. As a~result, contributions from
various multiphonon processes to the formation rate plotted
in~Fig.~\ref{fig:dtm_hd3k} cannot be distinguished. The resonance peak
for $\nu_f=2$ in HD is the strongest $dt\mu$ resonance found for the
three considered molecules.
\begin{figure}[htb] 
  \begin{center}
    \includegraphics[width=7cm]{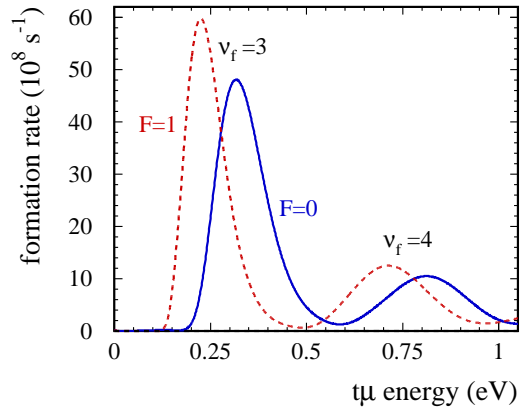}
    \caption{Resonant $dt\mu$-formation rate in 3-K solid~HT\@. 
      \label{fig:dtm_dt3k}}
  \end{center}
\end{figure}
In~Fig.~\ref{fig:dtm_dt3k}, the $dt\mu$-formation rate is shown for
a~3-K solid DT~target. The lowest peaks, which already take the
asymptotic form~(\ref{eq:res_asym_gam}), correspond here to~$\nu_f=3$.
The rotational and vibrational quanta are smallest for~DT, so that the
main (lowest $K_f$) resonances connected with~$\nu_f=2$ are located
deeply below~$\varepsilon=0$. Thus, a~contribution to the formation rate
from the subthreshold resonances is very small and is not apparent in
this figure. At~3~K, the effective target temperature~$T_\text{eff}$,
determined by~Eqs.~(\ref{eq:T_eff}) and~(\ref{eq:meankin}), equals about
41~K for~HD and 50~K for~DT\@. The resonance
shift~$\varDelta\varepsilon_{if}$ obtained from
Eq.~(\ref{eq:res_shift}), equals~$-2.71$~meV in the case of~HD and
$-1.97$~meV for~DT.

The $dt\mu$ resonances in solid HD and D$_2$ were directly observed
at~TRIUMF~\cite{fuji99,fuji00,porc01,mars01} using the energetic
$t\mu$-atom beam and time-of-flight techniques. However, Monte Carlo
simulations~(see e.g., Ref.~\cite{hube99}) were employed for
interpretation of the experimental data. Such a~procedure was
indispensable since the time-of-flight spectra cannot be uniquely
inverted because of the geometry used and the energy loss of $t\mu$
atoms in the reaction layer, prior to resonant formation of the
muonic-molecular complex. In those simulations, the calculated
$dt\mu$-formation rates for 3-K gas targets (such as that shown
in~Fig.~\ref{fig:dtm_d2gas3k}) were applied because the theoretical
formation rates for a~low-pressure solid were not available. A~detailed
analysis of the data was performed by Fujiwara~\cite{fuji99}. He found
more $dt$-fusion events at lowest and highest $t\mu$ energies than it
was predicted using the perfect gas model. Much broader resonance peaks,
which we present in Fig.~\ref{fig:dtm_nd2_3k}, can certainly improve the
fits to the TRIUMF data. Also, the analysis of the fusion-product
yield~\cite{fuji99} proved that the low-energy $dt\mu$-formation rate in
solid deuterium was much higher than that predicted by the two-collision
gas model. In particular, this concerns formation for the state~$F=1$.
The theoretical rates presented in~Fig.~\ref{fig:dtm_d2_sh3kf1} support
this finding.

A~two-peak structure of the calculated time-of-flight spectra for
$dt\mu$ resonances in~HD, obtained assuming a~3-K gas model, was not
confirmed by the solid-HD data~\cite{porc01}. However, one may expect
much better agreement with the TRIUMF data when the rate shown
in~Fig.~\ref{fig:dtm_hd3k} is used instead of the very pronounced peaks
evaluated for a~3-K HD gas. A~possibility of wider resonance peaks with
a~fixed Doppler width of~50~meV was already considered
in~Ref.~\cite{porc01}, which did not give good fits to the data.  Such
a~result is now clear since, according to~Eq.~(\ref{eq:Dopp_res}), the
Doppler width of a~resonance in a~condensed target increases with the
rising recoil energy~$\omega_\text{R}$. Simultaneously, the resonance
height~(\ref{eq:res_asym_gam}) decreases for higher~$\omega_\text{R}$
so that the resonance strength is preserved.

\begin{figure}[htb]
  \begin{center}
    \includegraphics[width=7cm]{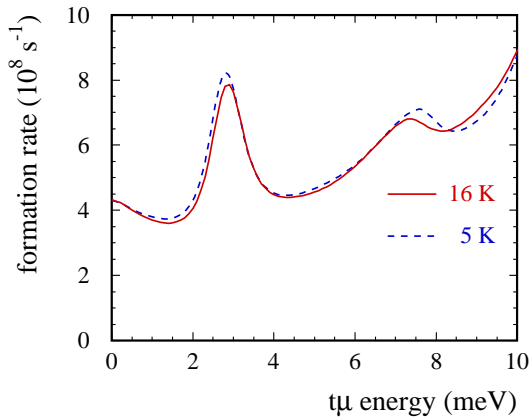}
    \caption{Rate of resonant $dt\mu$ formation in $t\mu(F=0)$ 
      scattering from a~D$_2$ molecule bound in 5-K and 16-K solid
      D/T($C_t=0.4$) targets.
    \label{fig:dtm_d2temp}}
  \end{center}
\end{figure}
\begin{figure}[htb]
  \begin{center} 
    \includegraphics[width=7cm]{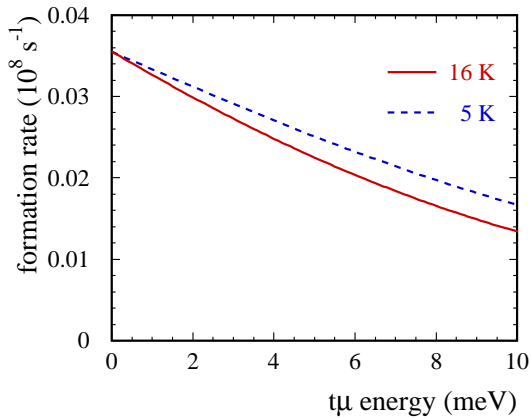}
    \caption{Rate of resonant $dt\mu$ formation in $t\mu(F=0)$ 
      scattering from a~DT molecule bound in 5-K and 16-K solid
      D/T($C_t=0.4$) targets.
    \label{fig:dtm_dttemp}}
  \end{center}
\end{figure}
In~Figs.~\ref{fig:dtm_d2temp} and~\ref{fig:dtm_dttemp} are shown the
resonant-formation rates for the molecules D$_2$ and~DT bound in a~solid
D/T target. An equilibrated mixture of the molecules D$_2$, DT,
and~T$_2$ is assumed, for the tritium isotopic concentration $C_t=0.4$.
Target temperatures 5--16~K were applied in the RIKEN-RAL experiment, in
which an unexpected temperature dependence of the $dt\mu$-formation rate
in solid D/T mixtures~\cite{kawa03} was found. The corresponding target
density is almost constant. A~similar hydrogens mixture, kept at 15~K,
was also used in the PSI experiment~\cite{acke99}. In the both
experiments, time spectra of neutrons from $dt$ fusion were measured.
The data were interpreted using a~standard steady-state kinetics,
assuming that $t\mu$ atoms were thermalized. Formation from the state
$F=1$ is negligible for an appreciable tritium concentration as the
spin-flip transition $F=1\to{}0$ in low-energy $t\mu+t$ collision is
very fast~\cite{adam96a}. The theoretical energy-dependent
$dt\mu$-formation rates display a~weak temperature dependence. One can
expect such a~behavior since, for any temperature of a~low-pressure
hydrogen-isotope solid, the limit $T/\varTheta_\text{D}\ll{}1$ is
achieved ($\varTheta_\text{D}\approx{}100$~K) and changes
of~$\varTheta_\text{D}$ are very small~\cite{silv80,soue86}. As
a~result, the response function~(\ref{eq:S_inter2}) and thus the
formation rate~(\ref{eq:resrat_inter}) are always close to their limits
for $T/\varTheta_\text{D}\to{}0$. Therefore, changes of the average
formation rate, determined using steady-state conditions, can only be
ascribed to different $t\mu$-energy distributions corresponding to
various target temperatures. An accurate comparison of the theory with
data requires Monte Carlo simulations of the $\mu$CF cycle in a~given
solid D/T mixture, which can be performed in future after completion of
a~full set of the differential cross section for muonic atom scattering
in mixed D/T crystals. The $t\mu$-energy distribution in steady-state
conditions is a~crucial information. A~shape of such a~distribution is
non-Maxwellian and the mean $t\mu$ energy is greater than
$\tfrac{3}{2}k_\text{B}{}T$, due to solid-state effects and a~possible
admixture of epithermal $t\mu$'s from the reaction $d\mu+t\to{}t\mu+d$
and from back decay of the muonic-molecular complex. The latter effect
was studied in~Refs.~\cite{cohe86,jeit95} with the use of Monte Carlo
simulations, in the case of gas and liquid targets. In a~high-density
target with medium or high~$C_t$, this effect is small, which is
confirmed by the PSI fits~\cite{acke99}.

\begin{figure}[htb] 
  \begin{center}
    \includegraphics[width=7cm]{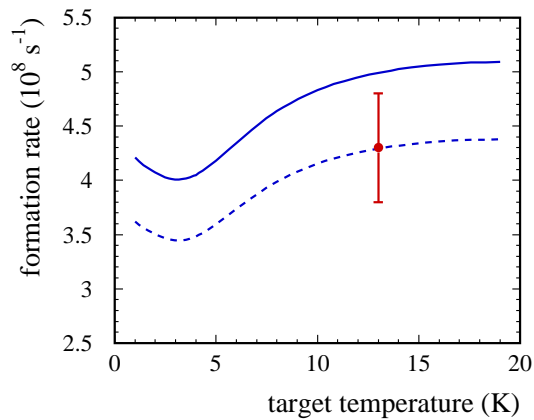}
    \caption{Mean rate of resonant $dt\mu$ formation in $t\mu(F=0)$
      scattering from nD$_2$ molecules bound in solid D/T ($C_t=0.4$) as
      a~function of the target temperature. The dashed line represents
      the same rate scaled by the factor $S_\lambda=0.86$. Also is shown
      the result of PSI measurement~\cite{acke99} for a~similar target
      ($T=13$~K, $\varphi=1.45$).
      \label{fig:dtm_d2_tavg}}
  \end{center}
\end{figure}

Averaging the energy-dependent rate from Fig.~\ref{fig:dtm_d2temp} over
the $t\mu$-energy distribution leads to the mean resonant rate shown
in~Fig.~\ref{fig:dtm_d2_tavg}.
The energy distribution of $t\mu$ atoms, being in thermal equilibrium
with phonons, is assumed to be proportional to
$Z(\varepsilon)\,{}n_\text{B}(\varepsilon,T)$. The average $t\mu$ energy
obtained using this function ranges from 1.2~meV for $T=5$~K to 3.4~meV
for $T=16$~K. It is evident that the rise of the formation rate, above
about~3~K, is mainly due to $t\mu$ entering into the region of the
recoil-less resonant peak in para-D$_2$, centered at~2.7~meV.  Phonon
processes in both ortho-D$_2$ and para-D$_2$ lead to a~smaller rise of
the rate. The calculated formation rate is close to the PSI result for
$T=13$~K~\cite{acke99}. A~coincidence of the theoretical curve with the
data is obtained, as in the case of the TRIUMF
measurements~\cite{fuji00}, on scaling by the factor $S_\lambda<1$,
which can be ascribed to inaccuracy of the calculated transition-matrix
elements. Here, we have $S_\lambda=0.86$, which is consistent with the
result of~Ref.~\cite{fuji00}.

In the RIKEN-RAL experiment~\cite{kawa03}, about 20\% decrease of the
$\mu$CF effectiveness has been found for the target-temperature change
from 16~to~5~K, independently of the tritium concentration. In order to
explain this effect, several hypotheses have been considered. The
hypothesis of a~significant change of the mean resonant $dt\mu$
formation rate~$\tilde{\lambda}_{dt\mu}^0$ (for $F=0$) has led to best
fits to the data. Kawamura~et~al assume that the two components
of~$\tilde{\lambda}_{dt\mu}^0$, namely the
rate~$\tilde{\lambda}_{dt\mu}^{0,\text{D}_2}$ of resonant formation for
the D$_2$~molecule and the analogous
rate~$\tilde{\lambda}_{dt\mu}^{0,\text{DT}}$ for the DT~molecule, are
comparable. At~16~K, they use
$\tilde{\lambda}_{dt\mu}^{0,\text{D}_2}=3.5\times{}10^8$~s$^{-1}$ and
$\tilde{\lambda}_{dt\mu}^{0,\text{DT}}=1.6\times{}10^8$~s$^{-1}$
\cite{kawa03}. All temperature dependence of
$\tilde{\lambda}_{dt\mu}^0\equiv{}%
C_d\,\tilde{\lambda}_{dt\mu}^{0,\text{D}_2}%
+C_t\,\tilde{\lambda}_{dt\mu}^{0,\text{DT}}$ ($C_d$ is the deuterium
isotopic concentration) is ascribed only
to~$\tilde{\lambda}_{dt\mu}^{0,\text{D}_2}$. Other rates in the
steady-state kinetics being fixed, about 30\% decrease
of~$\tilde{\lambda}_{dt\mu}^{0,\text{D}_2}$ between 16~K and~5~K has
been obtained. Thus, for $C_t=0.4$, the respective change
of~$\tilde{\lambda}_{dt\mu}^0$ equals about~25\%. This finding agrees
quite well with analogous 20\% decrease of the theoretical rate plotted
in~Fig.~\ref{fig:dtm_d2_tavg}. However, theory predicts that the
low-energy rate~$\tilde{\lambda}_{dt\mu}^{0,\text{DT}}$ should be
smaller by a~few orders of magnitude than the corresponding
rate~$\tilde{\lambda}_{dt\mu}^{0,\text{D}_2}$, since the strong
resonances in $t\mu+$DT scattering are far from the
region~$\varepsilon\approx{}0$. Averaging the rate presented
in~Fig.~\ref{fig:dtm_dttemp} over the $t\mu$-energy distribution gives
$\tilde{\lambda}_{dt\mu}^{0,\text{DT}}=2.6\times{}10^6$~s$^{-1}$. This
value agrees well with the rate
$\tilde{\lambda}_{dt\mu}^{0,DT}=(1.8\pm{}0.7)\times{}10^6$~s$^{-1}$,
determined for a~30\nobreakdash-K liquid D/T in the PSI
experiment~\cite{acke99}. Note that the formation rate in the solid is
somewhat greater than the corresponding rate in the liquid, which is
a~general law confirmed by experiments. Thus, according to the presented
calculation and to the PSI results,
$\tilde{\lambda}_{dt\mu}^0\approx\tilde{\lambda}_{dt\mu}^{0,\text{D}_2}$.
This means that in the steady-state analysis of Ref.~\cite{kawa03},
a~somewhat greater value of~$\tilde{\lambda}_{dt\mu}^{0,\text{D}_2}$
should have been assumed. In fact, Monte-Carlo simulations similar to
that performed for gaseous ~D/T~\cite{jeit95} are indispensable for an
accurate analysis of such experiments, since several rates change
significantly at lowest energies and thermalization process of muonic
atoms in solid hydrogens is complicated~\cite{adam99,adam01,wozn03}. It
depends on the target temperature, isotopic concentration, and
rotational population. A~full set of the differential cross sections for
muonic atom scattering in mixed solid D/T is necessary for accurate
description of $\mu$CF in such a~target.

\section{Conclusions}

A~method of calculating the rates of muonic-molecule resonant formation
in collision of muonic atoms with condensed hydrogens has been
developed. In the case of polycrystalline hydrogen-isotope targets,
detailed calculations have been performed using the Debye model of an
isotropic harmonic solid. Values of the resonant-formation rates have
been computed for resonant $dt\mu$ formation in frozen D/T and HD
targets, for collision energies $\lesssim{}1$~eV. These rates are very
different from those obtained for dilute gaseous hydrogens and exhibit
strong solid-state effects.

At lowest energies, contributions to the total rate from formation in
a~rigid lattice and from formation with simultaneous phonon processes
can be distinguished. In the high-energy region
($\varepsilon\gtrsim{}0.1$~eV), for any target, the rate takes a~general
asymptotic form, which depends on the mean kinetic energy of a~target
molecule. For low-pressure solid and liquid hydrogens, this energy is
much greater than the corresponding energy in a~perfect gas. As
a~result, condensed-matter effects in resonant formation do not
disappear even at highest collision energies. Since the main $dt\mu$
resonances for HD and DT are located far from zero energy, in these
cases it is sufficient to use only the asymptotic expression for the
formation rate.

The calculated resonance profiles in solid are much broader than in the
dilute-gas case. Experimental evidence supporting this conclusion has
been found in the time-of-flight measurements of $dt\mu$ resonances at
TRIUMF. A~quantitative comparison of the theory with these experiments
requires however complicated Monte-Carlo simulations.

The mean values of the $dt\mu$-formation rates for D$_2$ bound in the
solid D/T mixtures, averaged over the $t\mu$ kinetic energy under the
steady-state conditions, agree well with the PSI and RIKEN-RAL data.
Also a~temperature dependence of the mean formation rate, determined at
RIKEN-RAL for temperatures 5--16~K, is revealed by the theory.

\begin{acknowledgments}
  We would like to thank Profs. K.~Nagamine and L.~I.~Ponomarev for
  supporting this research and Drs. M.~C.~Fujiwara and G.~M.~Marshall
  for valuable discussions.
\end{acknowledgments}


\bibliography{adamczak}

\end{document}